\documentclass[aps,prb,amsmath,amssymb,10pt,a4paper,floatfix,twocolumn,superscriptaddress]{revtex4-1}

\pdfoutput=1

\usepackage{braket}
\usepackage{graphicx}
\usepackage[english]{babel}
\usepackage{color}
\usepackage{tabularx}
\usepackage{algorithm}
\usepackage{algpseudocode}
\usepackage{bm}
\usepackage{tikz}
\usepackage{subfigure}
\usepackage{verbatim} 
\usepackage{listings}
\usepackage[pdfpagelabels,plainpages=false,bookmarks=true,colorlinks,linkcolor=red,urlcolor=blue,citecolor=blue]{hyperref}   

\def\figpath{.}

\newcommand{\mpsUpFour}[2]{
\draw (#1-2.0,#2) node (X) {$\dots$};
\AU{#1}{#2}
\AU{#1+2.0}{#2}
\AU{#1+4.0}{#2}
\AU{#1+6.0}{#2}
\draw (#1+8.0,#2) node (X) {$\dots$};
}

\newcommand{\mpsUpFive}[2]{
\draw (#1-2.0,#2) node (X) {$\dots$};
\AU{#1}{#2}
\AU{#1+2.0}{#2}
\AU{#1+4.0}{#2}
\AU{#1+6.0}{#2}
\AU{#1+8.0}{#2}
\draw (#1+10.0,#2) node (X) {$\dots$};
}

\newcommand{\mpsDownFive}[2]{
\draw (#1-2.0,#2) node (X) {$\dots$};
\AD{#1}{#2}
\AD{#1+2.0}{#2}
\AD{#1+4.0}{#2}
\AD{#1+6.0}{#2}
\AD{#1+8.0}{#2}
\draw (#1+10.0,#2) node (X) {$\dots$};
}

\newcommand{\mpoOneFour}[2]{
\T{#1}{#2}
\T{#1+2.0}{#2}
\T{#1+4.0}{#2}
\T{#1+6.0}{#2}
}

\newcommand{\mpoFourOne}[2]{
\T{#1}{#2}
\T{#1}{#2+2.0}
\T{#1}{#2+4.0}
\T{#1}{#2+6.0}
}

\newcommand{\rowtorow}[2]{
\draw (#1-2.0,#2) node (X) {$\dots$};
\mpoOneFour{#1}{#2}
\draw (#1+8.0,#2) node (X) {$\dots$};
}

\newcommand{\rowtorowFive}[2]{
\draw (#1-2.0,#2) node (X) {$\dots$};
\mpoOneFive{#1}{#2}
\draw (#1+10.0,#2) node (X) {$\dots$};
}

\newcommand{\rowtorowFiveX}[2]{
\draw (#1-2.0,#2) node (X) {$\dots$};
\mpoOneFiveXThree{#1}{#2}
\draw (#1+10.0,#2) node (X) {$\dots$};
}

\newcommand{\coltocol}[2]{
\draw (#1,#2-2.0) node (X) {$\vdots$};
\mpoFourOne{#1}{#2}
\draw (#1,#2+8.0) node (X) {$\vdots$};
}

\newcommand{\mpoOneFive}[2]{
\T{#1}{#2}
\T{#1+2.0}{#2}
\T{#1+4.0}{#2}
\T{#1+6.0}{#2}
\T{#1+8.0}{#2}
}

\newcommand{\mpoOneFourX}[2]{
\TX{#1}{#2}
\T{#1+2.0}{#2}
\T{#1+4.0}{#2}
\T{#1+6.0}{#2}
}

\newcommand{\mpoOneFiveXThree}[2]{
\T{#1}{#2}
\T{#1+2.0}{#2}
\TX{#1+4.0}{#2}
\T{#1+6.0}{#2}
\T{#1+8.0}{#2}
}

\newcommand{\mpoOneFourY}[2]{
\T{#1}{#2}
\T{#1+2.0}{#2}
\T{#1+4.0}{#2}
\TY{#1+6.0}{#2}
}

\newcommand{\mpoFourFour}[2]{
\mpoOneFour{#1}{#2}
\mpoOneFour{#1}{#2+2.0}
\mpoOneFour{#1}{#2+4.0}
\mpoOneFour{#1}{#2+6.0}
}

\newcommand{\mpoFourFourXY}[2]{
\mpoOneFourX{#1}{#2}
\mpoOneFour{#1}{#2+2.0}
\mpoOneFourY{#1}{#2+4.0}
\mpoOneFour{#1}{#2+6.0}
}

\newcommand{\mpsFiveMixed}[2]{
\MPSuLtens{#1}{#2}{A_U^L}
\MPSuLtens{#1+2.0}{#2}{A_U^L}
\MPSuCtens{#1+4.0}{#2}{A_U^C}
\MPSuRtens{#1+6.0}{#2}{A_U^R}
\MPSuRtens{#1+8.0}{#2}{A_U^R}
}

\newcommand{\leftbracket}[3]{
\lineV{#2}{#2+#3}{#1}
\lineH{#1}{#1+0.25}{#2+#3}
\lineH{#1}{#1+0.25}{#2}
}

\newcommand{\rightbracket}[3]{
\lineV{#2}{#2+#3}{#1}
\lineH{#1-0.25}{#1}{#2+#3}
\lineH{#1-0.25}{#1}{#2}
}

\newcommand{\dotsFourOne}[2]{
\draw (#1,#2) node (X) {$\dots$};
\draw (#1,#2+2.0) node (X) {$\dots$};
\draw (#1,#2+4.0) node (X) {$\dots$};
\draw (#1,#2+6.0) node (X) {$\dots$};
}

\newcommand{\dotsOneFour}[2]{
\draw (#1,#2) node (X) {$\vdots$};
\draw (#1+2.0,#2) node (X) {$\vdots$};
\draw (#1+4.0,#2) node (X) {$\vdots$};
\draw (#1+6.0,#2) node (X) {$\vdots$};
}

\newcommand{\tracetwodimnetwork}[2]{
\draw[scale=1.0] (#1-5.0,#2+3.0) node (X) {$\kappa^{MN} \equiv Tr$};
\dotsOneFour{#1}{#2-2.0}
\leftbracket{#1-3.0}{#2-2.75}{11.0}
\dotsFourOne{#1-2.0}{#2}
\mpoFourFour{#1}{#2}
\dotsFourOne{#1+8.0}{#2}
\dotsOneFour{#1}{#2+8.0}
\rightbracket{#1+9.0}{#2-2.75}{11.0}
}

\newcommand{\tracetwodimnetworkXY}[2]{
\draw[scale=1.0] (#1-5.0,#1+3.0) node (X) {$\langle XY\rangle = Tr$};
\leftbracket{#1-3.0}{#2-2.75}{11.0}
\dotsOneFour{#1}{#2-2.0}
\dotsFourOne{#1-2.0}{#2}
\mpoFourFourXY{#1}{#2}
\dotsFourOne{#1+8.0}{#2}
\dotsOneFour{#1}{#2+8.0}
\rightbracket{#1+9.0}{#2-2.75}{11.0}
\draw[scale=1.0] (#1+10.5,#2+3.0) node (X) {$/\ \kappa^{MN}$};
}

\newcommand{\ctmrgansatzTwoTwo}[2]{
\CULT{#1}{#2}
\CURT{#1+2.0}{#2}
\CDRT{#1+2.0}{#2-2.0}
\CDLT{#1}{#2-2.0}
}

\newcommand{\ctmrgansatzTwoTwoSym}[2]{
\CULTsym{#1}{#2}
\CURTsym{#1+2.0}{#2}
\CDRTsym{#1+2.0}{#2-2.0}
\CDLTsym{#1}{#2-2.0}
}

\newcommand{\T}[2]{
\MPOtens{#1}{#2}{T}
}

\newcommand{\TX}[2]{
\MPOtens{#1}{#2}{T_X}
}

\newcommand{\TY}[2]{
\MPOtens{#1}{#2}{T_Y}
}

\newcommand{\AU}[2]{
\MPSutens{#1}{#2}{A_U}
}

\newcommand{\AD}[2]{
\MPSdtens{#1}{#2}{A_D}
}

\newcommand{\ALp}[2]{
\MPSltens{#1}{#2}{A_L'}
}

\newcommand{\ALpsym}[2]{
\MPSltens{#1}{#2}{A'}
}

\newcommand{\ALT}[2]{
\MPSltens{#1-2.0}{#2}{A_L}
\MPOtens{#1}{#2}{T}
}

\newcommand{\ALTsym}[2]{
\MPSltens{#1-2.0}{#2}{A}
\MPOtens{#1}{#2}{T}
}

\newcommand{\ALpT}[2]{
\MPSltens{#1-2.0}{#2}{A_L'}
\MPOtens{#1}{#2}{T}
}

\newcommand{\CUCDrp}[2]{
\centerCudptens{#1+2.5}{#2+1.5}{C_U^{(1)}}
\cornerIulptens{#1}{#2+1.5}
\cornerIdlptens{#1}{#2-1.5}
\centerCudptens{#1+2.5}{#2-1.5}{C_D^{(1)}}
}

\newcommand{\svdCUCDrp}[2]{
\cornerIultens{#1}{#2+1.5}
\centerClrtens{#1}{#2}{\Sigma_L^2}
\cornerIdltens{#1}{#2-1.5}
\mpsARp{#1+1.5}{#2+1.5}{\bar{V}_L}
\mpsARp{#1+1.5}{#2-1.5}{U_L}
}

\newcommand{\CudOneOne}[2]{
\CUOneOne{#1}{#2}
\CDOneOne{#1}{#2-4.0}
}

\newcommand{\CULp}[2]{
\cornerCultens{#1}{#2}{C_{LU}'}
}

\newcommand{\CUOneOne}[2]{
\centerCudptens{#1}{#2}{C_U^{(1)}}
\cornerIulptens{#1-2.5}{#2};
\cornerIurptens{#1+2.5}{#2};
}

\newcommand{\CDOneOne}[2]{
\centerCudptens{#1}{#2}{C_D^{(1)}}
\cornerIdlptens{#1-2.5}{#2};
\cornerIdrptens{#1+2.5}{#2};
}

\newcommand{\CULT}[2]{
\cornerCultens{#1-2.0}{#2+2.0}{C_{LU}}
\MPSutens{#1}{#2+2.0}{A_U}
\MPSltens{#1-2.0}{#2}{A_L}
\MPOtens{#1}{#2}{T}
}

\newcommand{\CULTsym}[2]{
\cornerCultens{#1-2.0}{#2+2.0}{C}
\MPSutens{#1}{#2+2.0}{A}
\MPSltens{#1-2.0}{#2}{A}
\MPOtens{#1}{#2}{T}
}

\newcommand{\CULTp}[2]{
\cornerCultens{#1-2.0}{#2+2.0}{C_{LU}'}
\MPSutens{#1}{#2+2.0}{A_U}
\MPSltens{#1-2.0}{#2}{A_L}
\MPOtens{#1}{#2}{T}
}

\newcommand{\CULAU}[2]{
\cornerCultens{#1-2.0}{#2}{C_{LU}}
\MPSutens{#1}{#2}{A_U}
}

\newcommand{\CULAUrenorm}[2]{
\CULAU{#1}{#2}
\PLp{#1}{#2-2.0}
\Ilrtens{#1+2.0}{#2}
}

\newcommand{\CULsymeig}[2]{
\MPSuLtens{#1-2.0}{#2}{U}
\cornerIultens{#1-4.0}{#2}
\cornerCdrtens{#1}{#2}{D}
\MPSlUtens{#1}{#2+2.0}{\bar{U}}
\cornerIultens{#1}{#2+4.0}
}

\newcommand{\CURp}[2]{
\cornerCurtens{#1}{#2}{C_{UR}'}
}

\newcommand{\CURT}[2]{
\cornerCurtens{#1+2.0}{#2+2.0}{C_{UR}}
\MPSutens{#1}{#2+2.0}{A_U}
\MPSrtens{#1+2.0}{#2}{A_R}
\MPOtens{#1}{#2}{T}
}

\newcommand{\CURTsym}[2]{
\cornerCurtens{#1+2.0}{#2+2.0}{C}
\MPSutens{#1}{#2+2.0}{A}
\MPSrtens{#1+2.0}{#2}{A}
\MPOtens{#1}{#2}{T}
}

\newcommand{\CURTp}[2]{
\cornerCurtens{#1+2.0}{#2+2.0}{C_{UR}'}
\MPSutens{#1}{#2+2.0}{A_U}
\MPSrtens{#1+2.0}{#2}{A_R}
\MPOtens{#1}{#2}{T}
}

\newcommand{\CDRp}[2]{
\cornerCdrtens{#1}{#2}{C_{RD}'}
}

\newcommand{\CDRT}[2]{
\cornerCdrtens{#1+2.0}{#2-2.0}{C_{RD}}
\MPSdtens{#1}{#2-2.0}{A_D}
\MPSrtens{#1+2.0}{#2}{A_R}
\MPOtens{#1}{#2}{T}
}

\newcommand{\CDRTsym}[2]{
\cornerCdrtens{#1+2.0}{#2-2.0}{C}
\MPSdtens{#1}{#2-2.0}{A}
\MPSrtens{#1+2.0}{#2}{A}
\MPOtens{#1}{#2}{T}
}

\newcommand{\CDRTp}[2]{
\cornerCdrtens{#1+2.0}{#2-2.0}{C_{RD}'}
\MPSdtens{#1}{#2-2.0}{A_D}
\MPSrtens{#1+2.0}{#2}{A_R}
\MPOtens{#1}{#2}{T}
}

\newcommand{\CDLp}[2]{
\cornerCdltens{#1}{#2}{C_{DL}'}
}

\newcommand{\CDLT}[2]{
\cornerCdltens{#1-2.0}{#2-2.0}{C_{DL}}
\MPSdtens{#1}{#2-2.0}{A_D}
\MPSltens{#1-2.0}{#2}{A_L}
\MPOtens{#1}{#2}{T}
}

\newcommand{\CDLTsym}[2]{
\cornerCdltens{#1-2.0}{#2-2.0}{C}
\MPSdtens{#1}{#2-2.0}{A}
\MPSltens{#1-2.0}{#2}{A}
\MPOtens{#1}{#2}{T}
}

\newcommand{\CDLTp}[2]{
\cornerCdltens{#1-2.0}{#2-2.0}{C_{DL}'}
\MPSdtens{#1}{#2-2.0}{A_D}
\MPSltens{#1-2.0}{#2}{A_L}
\MPOtens{#1}{#2}{T}
}

\newcommand{\CDLAD}[2]{
\cornerCdltens{#1-2.0}{#2}{C_{DL}}
\MPSdtens{#1}{#2}{A_D}
}

\newcommand{\CDLADrenorm}[2]{
\CDLAD{#1}{#2}
\PL{#1}{#2+2.0}
\Ilrtens{#1+2.0}{#2}
}

\newcommand{\SVDU}[2]{
\centerCudtens{#1}{#2}{S_U}
\MPSuLtens{#1-2.0}{#2}{U_U}
\MPSuRtens{#1+2.0}{#2}{\bar{V}_U}
\cornerIultens{#1-4.0}{#2}
\cornerIurtens{#1+4.0}{#2}
}

\newcommand{\SVDD}[2]{
\centerCudtens{#1}{#2}{S_D}
\MPSdLtens{#1-2.0}{#2}{\bar{V}_D}
\MPSdRtens{#1+2.0}{#2}{U_D}
\cornerIdltens{#1-4.0}{#2}
\cornerIdrtens{#1+4.0}{#2}
}

\newcommand{\factorU}[2]{
\MPSuCtens{#1-1.0}{#2}{F_{LU}}
\MPSuCtens{#1+1.0}{#2}{F_{UR}}
\cornerIultens{#1-3.0}{#2}
\cornerIurtens{#1+3.0}{#2}
}

\newcommand{\factorD}[2]{
\MPSdCtens{#1-1.0}{#2}{F_{DL}}
\MPSdCtens{#1+1.0}{#2}{F_{RD}}
\cornerIdltens{#1-3.0}{#2}
\cornerIdrtens{#1+3.0}{#2}
}

\newcommand{\QRU}[2]{
\MPSuCtens{#1-1.0}{#2}{R_U^T}
\MPSuRtens{#1+1.0}{#2}{Q_U^T}
\cornerIultens{#1-3.0}{#2}
\cornerIurtens{#1+3.0}{#2}
}

\newcommand{\QRD}[2]{
\MPSdCtens{#1-1.0}{#2}{R_D}
\MPSdRtens{#1+1.0}{#2}{Q_D}
\cornerIdltens{#1-3.0}{#2}
\cornerIdrtens{#1+3.0}{#2}
}

\newcommand{\PU}[2]{
\MPSlUtens{#1}{#2}{P_U^-}
\cornerIdrtens{#1}{#2-2.0}
\cornerIultens{#1}{#2+2.0}
}

\newcommand{\PRp}[2]{
\MPSdRtens{#1}{#2}{P_R}
\cornerIdrtens{#1+2.0}{#2}
\cornerIultens{#1-2.0}{#2}
}

\newcommand{\PLp}[2]{
\MPSdLtens{#1}{#2}{P_L^-}
\cornerIdltens{#1-2.0}{#2}
\cornerIurtens{#1+2.0}{#2}
}

\newcommand{\PLpsym}[2]{
\MPSdLtens{#1}{#2}{\bar{U}}
\cornerIdltens{#1-2.0}{#2}
\cornerIurtens{#1+2.0}{#2}
}

\newcommand{\PUp}[2]{
\MPSrUtens{#1}{#2}{P_U}
\cornerIdltens{#1}{#2-2.0}
\cornerIurtens{#1}{#2+2.0}
}

\newcommand{\PUpsym}[2]{
\MPSrUtens{#1}{#2}{U}
\cornerIdltens{#1}{#2-2.0}
\cornerIurtens{#1}{#2+2.0}
}

\newcommand{\PR}[2]{
\MPSuRtens{#1}{#2}{P_R^-}
\cornerIurtens{#1+2.0}{#2}
\cornerIdltens{#1-2.0}{#2}
}

\newcommand{\PD}[2]{
\MPSlDtens{#1}{#2}{P_D}
\cornerIurtens{#1}{#2+2.0}
\cornerIdltens{#1}{#2-2.0}
}

\newcommand{\PL}[2]{
\MPSuLtens{#1}{#2}{P_L}
\cornerIultens{#1-2.0}{#2}
\cornerIdrtens{#1+2.0}{#2}
}

\newcommand{\PLsym}[2]{
\MPSuLtens{#1}{#2}{U}
\cornerIultens{#1-2.0}{#2}
\cornerIdrtens{#1+2.0}{#2}
}

\newcommand{\PDp}[2]{
\MPSrDtens{#1}{#2}{P_D^-}
\cornerIultens{#1}{#2+2.0}
\cornerIdrtens{#1}{#2-2.0}
}

\newcommand{\PUPL}[2]{
\PLp{#1}{#2-2.0}
\PUp{#1+2.0}{#2}
}

\newcommand{\PUPLsym}[2]{
\PLpsym{#1}{#2-2.0}
\PUpsym{#1+2.0}{#2}
}

\newcommand{\PUPR}[2]{
\PU{#1-2.0}{#2}
\PRp{#1}{#2-2.0}
}

\newcommand{\PDPR}[2]{
\PD{#1-2.0}{#2}
\PR{#1}{#2+2.0}
}

\newcommand{\PDPL}[2]{
\PDp{#1+2.0}{#2}
\PL{#1}{#2+2.0}
}

\newcommand{\PLPL}[2]{
\PL{#1}{#2+2.0}
\PLp{#1}{#2-2.0}
}

\newcommand{\PLPLsym}[2]{
\PLsym{#1}{#2+2.0}
\PLpsym{#1}{#2-2.0}
}

\newcommand{\CULTrenormsym}[2]{
\CULTsym{#1}{#2}
\PUPLsym{#1}{#2}
}

\newcommand{\CULTprenorm}[2]{
\CULTp{#1}{#2}
\PUPL{#1}{#2}
}

\newcommand{\CURTprenorm}[2]{
\CURTp{#1}{#2}
\PUPR{#1}{#2}
}

\newcommand{\CDRTprenorm}[2]{
\CDRTp{#1}{#2}
\PDPR{#1}{#2}
}

\newcommand{\CDLTprenorm}[2]{
\CDLTp{#1}{#2}
\PDPL{#1}{#2}
}

\newcommand{\ALrenorm}[2]{
\ALT{#1}{#2}
\PLPL{#1}{#2}
\Ilrtens{#1+2.0}{#2}
}

\newcommand{\ALprenorm}[2]{
\ALpT{#1}{#2}
\PLPL{#1}{#2}
\Ilrtens{#1+2.0}{#2}
}

\newcommand{\ALrenormsym}[2]{
\ALTsym{#1}{#2}
\PLPLsym{#1}{#2}
\Ilrtens{#1+2.0}{#2}
}

\newcommand{\MPStransferC}[4]{
\MPSuCtens{#1}{#2}{#3}
\MPSdCtens{#1}{#2-2.0}{#4}
}

\newcommand{\defPL}[2]{
\PL{#1}{#2+1.0}
\PLp{#1}{#2-1.0}
\draw (#1+3.0,#2) node {$\approx$};
\Iudtens{#1+4}{#2+1.0}
\Iudtens{#1+4}{#2-1.0}
}

\newcommand{\defprojL}[2]{
\PLp{#1}{#2+1.0}
\PL{#1}{#2-1.0}
}

\newcommand{\defL}[4]{
\cornerIultens{#1-2.0}{#2+1.0}
\cornerIdltens{#1-2.0}{#2-1.0}
\MPSuLtens{#1}{#2+1.0}{#3}
\MPSdLtens{#1}{#2-1.0}{#4}
\cornerIdrtens{#1+2.0}{#2+1.0}
\cornerIurtens{#1+2.0}{#2-1.0}
\draw (#1+3.0,#2) node {$=$};
\Iudtens{#1+4.0}{#2+1.0}
\Iudtens{#1+4.0}{#2-1.0}
}

\newcommand{\residualLHS}[2]{
\cornerIultens{#1}{#2}
\MPSuCtens{#1+2.0}{#2}{A_U^C}
\cornerIurtens{#1+4.0}{#2}
\MPSltens{#1}{#2-2.0}{E_L}
\MPOtens{#1+2.0}{#2-2.0}{T}
\MPSrtens{#1+4.0}{#2-2.0}{E_R}
\cornerIdrtens{#1}{#2-4.0}
\cornerIdltens{#1+4.0}{#2-4.0}
\Iudtens{#1+2.0}{#2-4.0}
}

\newcommand{\residualRHS}[2]{
\MPSuLtens{#1-2.0}{#2-4.0}{A_U^L}
\cornerIultens{#1}{#2}
\centerCudtens{#1+2.0}{#2}{C_U}
\cornerIurtens{#1+4.0}{#2}
\MPSltens{#1}{#2-2.0}{E_L}
\centerIudtens{#1+2.0}{#2-2.0}
\MPSrtens{#1+4.0}{#2-2.0}{E_R}
\cornerIdrtens{#1}{#2-4.0}
\cornerIdltens{#1+4.0}{#2-4.0}
}

\newcommand{\defLvumps}[4]{
\cornerIultens{#1-2.0}{#2+1.0}
\cornerIdltens{#1-2.0}{#2-1.0}
\MPSuLtens{#1}{#2+1.0}{#3}
\MPSdLtens{#1}{#2-1.0}{#4}
\draw (#1+2.0,#2) node {$=$};
\cornerIultens{#1+3.0}{#2+1.0}
\cornerIdltens{#1+3.0}{#2-1.0}
}

\newcommand{\defRvumps}[4]{
\cornerIurtens{#1}{#2+1.0}
\cornerIdrtens{#1}{#2-1.0}
\MPSuRtens{#1-2.0}{#2+1.0}{#3}
\MPSdRtens{#1-2.0}{#2-1.0}{#4}
\draw (#1+1.0,#2) node {$=$};
\cornerIurtens{#1+3.0}{#2+1.0}
\cornerIdrtens{#1+3.0}{#2-1.0}
}

\newcommand{\defprojLsym}[2]{
\PLpsym{#1}{#2+1.0}
\PLsym{#1}{#2-1.0}
}

\newcommand{\defPLsym}[2]{
\PLsym{#1}{#2+1.0}
\PLpsym{#1}{#2-1.0}
\draw (#1+3.0,#2) node {$=$};
\Iudtens{#1+4}{#2+1.0}
\Iudtens{#1+4}{#2-1.0}
}

\newcommand{\mpsA}[3]{\draw (#1,#2) circle (.5); \draw (#1,#2) node {$#3$};}

\newcommand{\mpsAL}[3]{\draw (#1+0.5,#2) --(#1+0.25,#2+0.5) --(#1-0.5,#2+0.5) --(#1-0.5,#2-0.5) --(#1+0.25,#2-0.5) --cycle; \draw (#1-0.05,#2) node {$#3$};}
\newcommand{\mpsAR}[3]{\draw (#1-0.5,#2) --(#1-0.25,#2+0.5) --(#1+0.5,#2+0.5) --(#1+0.5,#2-0.5) --(#1-0.25,#2-0.5) --cycle; \draw (#1+0.1,#2) node {$#3$};}
\newcommand{\mpsAU}[3]{\draw (#1,#2-0.5) --(#1+0.5,#2-0.25) --(#1+0.5,#2+0.5) --(#1-0.5,#2+0.5) --(#1-0.5,#2-0.25) --cycle; \draw (#1,#2) node {$#3$};}
\newcommand{\mpsAD}[3]{\draw (#1,#2+0.5) --(#1+0.5,#2+0.25) --(#1+0.5,#2-0.5) --(#1-0.5,#2-0.5) --(#1-0.5,#2+0.25) --cycle; \draw (#1,#2) node {$#3$};}
\newcommand{\mpsAC}[3]{\draw[rounded corners] (0.5+#1,0.5+#2) rectangle (-0.5+#1,-0.5+#2); \draw (#1,#2) node {$#3$};}
\newcommand{\mpoT}[3]{\draw (0.5+#1,0.5+#2) rectangle (-0.5+#1,-0.5+#2); \draw (#1,#2) node {$#3$};}
\newcommand{\centerC}[3]{\draw[rounded corners] (0.5+#1,0.5+#2) rectangle (-0.5+#1,-0.5+#2); \draw (#1,#2) node {$#3$};}
\newcommand{\cornerC}[3]{\draw[rounded corners] (0.5+#1,0.5+#2) rectangle (-0.5+#1,-0.5+#2); \draw (#1,#2) node {$#3$};}

\newcommand{\gaugeX}[3]{\draw (#1,#2) circle (.5); \draw (#1,#2) node {$#3$};}

\newcommand{\cornerCp}[3]{\draw[rounded corners] (1.0+#1,1.0+#2) rectangle (-1.0+#1,-1.0+#2); \draw (#1,#2) node {$#3$};}

\newcommand{\mpsALp}[3]{
\draw (#1+1.0,#2) --(#1+0.5,#2+1.0) --(#1-0.5,#2+1.0) --(#1-0.5,#2-1.0) --(#1+0.5,#2-1.0) --cycle;
\draw (#1+0.25,#2) node {$#3$};
\draw (#1-0.5,#2+0.5) -- (#1-1.0,#2+0.5);
\draw (#1-0.5,#2-0.5) -- (#1-1.0,#2-0.5);
}

\newcommand{\mpsARp}[3]{
\draw (#1-0.5,#2) --(#1,#2+1.0) --(#1+1.0,#2+1.0) --(#1+1.0,#2-1.0) --(#1,#2-1.0) --cycle; 
\draw (#1+0.4,#2) node {$#3$};
\draw (#1+1.0,#2+0.5) -- (#1+1.5,#2+0.5);
\draw (#1+1.0,#2-0.5) -- (#1+1.5,#2-0.5);
}

\newcommand{\lineH}[3]{\draw (#1,#3) -- (#2,#3);}
\newcommand{\lineV}[3]{\draw (#3,#1) -- (#3,#2);}

\newcommand{\lineCul}[2]{\draw[rounded corners] (#1,#2-0.5) --(#1,#2) --(#1+0.5,#2);}
\newcommand{\lineCdl}[2]{\draw[rounded corners] (#1,#2+0.5) --(#1,#2) --(#1+0.5,#2);}
\newcommand{\lineCur}[2]{\draw[rounded corners] (#1,#2-0.5) --(#1,#2) --(#1-0.5,#2);}
\newcommand{\lineCdr}[2]{\draw[rounded corners] (#1,#2+0.5) --(#1,#2) --(#1-0.5,#2);}

\newcommand{\cornerIultens}[2]{
\lineCul{#1}{#2};
\lineH{#1+0.5}{#1+1.0}{#2};
\lineV{#2-0.5}{#2-1.0}{#1};
}

\newcommand{\cornerIdltens}[2]{
\lineCdl{#1}{#2};
\lineH{#1+0.5}{#1+1.0}{#2};
\lineV{#2+0.5}{#2+1.0}{#1};
}

\newcommand{\cornerIurtens}[2]{
\lineCur{#1}{#2};
\lineH{#1-0.5}{#1-1.0}{#2};
\lineV{#2-0.5}{#2-1.0}{#1};
}

\newcommand{\cornerIdrtens}[2]{
\lineCdr{#1}{#2};
\lineH{#1-0.5}{#1-1.0}{#2};
\lineV{#2+0.5}{#2+1.0}{#1};
}

\newcommand{\Iudtens}[2]{
\lineV{#2-1.0}{#2+1.0}{#1};
}

\newcommand{\Ilrtens}[2]{
\lineH{#1-1.0}{#1+1.0}{#2};
}

\newcommand{\centerIudtens}[2]{
\lineH{#1-0.5}{#1+0.5}{#2};
\lineH{#1-0.5}{#1-1.0}{#2};
\lineH{#1+0.5}{#1+1.0}{#2};
}

\newcommand{\MPOtens}[3]{
\mpoT{#1}{#2}{#3};
\lineH{#1-0.5}{#1-1.0}{#2};
\lineH{#1+0.5}{#1+1.0}{#2};
\lineV{#2+0.5}{#2+1.0}{#1};
\lineV{#2-0.5}{#2-1.0}{#1};
}

\newcommand{\MPSutens}[3]{
\mpsA{#1}{#2}{#3};
\lineH{#1-0.5}{#1-1.0}{#2};
\lineH{#1+0.5}{#1+1.0}{#2};
\lineV{#2-0.5}{#2-1.0}{#1};
}

\newcommand{\MPSuLtens}[3]{
\mpsAL{#1}{#2}{#3};
\lineH{#1-1.0}{#1-0.5}{#2};
\lineH{#1+0.5}{#1+1.0}{#2};
\lineV{#2-1.0}{#2-0.5}{#1};
}

\newcommand{\MPSuRtens}[3]{
\mpsAR{#1}{#2}{#3};
\lineH{#1-0.5}{#1-1.0}{#2};
\lineH{#1+0.5}{#1+1.0}{#2};
\lineV{#2-0.5}{#2-1.0}{#1};
}

\newcommand{\MPSuCtens}[3]{
\mpsAC{#1}{#2}{#3};
\lineH{#1-0.5}{#1-1.0}{#2};
\lineH{#1+0.5}{#1+1.0}{#2};
\lineV{#2-0.5}{#2-1.0}{#1};
}

\newcommand{\MPSrtens}[3]{
\mpsA{#1}{#2}{#3};
\lineH{#1-0.5}{#1-1.0}{#2};
\lineV{#2-0.5}{#2-1.0}{#1};
\lineV{#2+0.5}{#2+1.0}{#1};
}

\newcommand{\MPSrUtens}[3]{
\mpsAU{#1}{#2}{#3};
\lineH{#1-0.5}{#1-1.0}{#2};
\lineV{#2-0.5}{#2-1.0}{#1};
\lineV{#2+0.5}{#2+1.0}{#1};
}

\newcommand{\MPSlUtens}[3]{
\mpsAU{#1}{#2}{#3};
\lineH{#1+0.5}{#1+1.0}{#2};
\lineV{#2-0.5}{#2-1.0}{#1};
\lineV{#2+0.5}{#2+1.0}{#1};
}

\newcommand{\MPSrDtens}[3]{
\mpsAD{#1}{#2}{#3};
\lineH{#1-0.5}{#1-1.0}{#2};
\lineV{#2-0.5}{#2-1.0}{#1};
\lineV{#2+0.5}{#2+1.0}{#1};
}

\newcommand{\MPSlDtens}[3]{
\mpsAD{#1}{#2}{#3};
\lineH{#1+0.5}{#1+1.0}{#2};
\lineV{#2-0.5}{#2-1.0}{#1};
\lineV{#2+0.5}{#2+1.0}{#1};
}

\newcommand{\MPSdtens}[3]{
\mpsA{#1}{#2}{#3};
\lineH{#1-0.5}{#1-1.0}{#2};
\lineH{#1+0.5}{#1+1.0}{#2};
\lineV{#2+0.5}{#2+1.0}{#1};
}

\newcommand{\MPSdLtens}[3]{
\mpsAL{#1}{#2}{#3};
\lineH{#1-0.5}{#1-1.0}{#2};
\lineH{#1+0.5}{#1+1.0}{#2};
\lineV{#2+0.5}{#2+1.0}{#1};
}

\newcommand{\MPSdRtens}[3]{
\mpsAR{#1}{#2}{#3};
\lineH{#1-0.5}{#1-1.0}{#2};
\lineH{#1+0.5}{#1+1.0}{#2};
\lineV{#2+0.5}{#2+1.0}{#1};
}

\newcommand{\MPSdCtens}[3]{
\mpsAC{#1}{#2}{#3};
\lineH{#1-0.5}{#1-1.0}{#2};
\lineH{#1+0.5}{#1+1.0}{#2};
\lineV{#2+0.5}{#2+1.0}{#1};
}

\newcommand{\MPSltens}[3]{
\mpsA{#1}{#2}{#3};
\lineH{#1+0.5}{#1+1.0}{#2};
\lineV{#2-0.5}{#2-1.0}{#1};
\lineV{#2+0.5}{#2+1.0}{#1};
}

\newcommand{\centerCudtens}[3]{
\centerC{#1}{#2}{#3};
\lineH{#1-0.5}{#1-1.0}{#2};
\lineH{#1+0.5}{#1+1.0}{#2};
}

\newcommand{\centerClrtens}[3]{
\centerC{#1}{#2}{#3};
\lineV{#2-0.5}{#2-1.0}{#1};
\lineV{#2+0.5}{#2+1.0}{#1};
}

\newcommand{\gaugeXudtens}[3]{
\gaugeX{#1}{#2}{#3};
\lineH{#1-0.5}{#1-1.0}{#2};
\lineH{#1+0.5}{#1+1.0}{#2};
}

\newcommand{\gaugeARudtens}[3]{
\mpsAR{#1}{#2}{#3};
\lineH{#1-0.5}{#1-1.0}{#2};
\lineH{#1+0.5}{#1+1.0}{#2};
}

\newcommand{\gaugeALudtens}[3]{
\mpsAL{#1}{#2}{#3};
\lineH{#1-0.5}{#1-1.0}{#2};
\lineH{#1+0.5}{#1+1.0}{#2};
}

\newcommand{\gaugeXlrtens}[3]{
\gaugeX{#1}{#2}{#3};
\lineV{#2-0.5}{#2-1.0}{#1};
\lineV{#2+0.5}{#2+1.0}{#1};
}

\newcommand{\cornerCultens}[3]{
\cornerC{#1}{#2}{#3};
\lineH{#1+0.5}{#1+1.0}{#2};
\lineV{#2-0.5}{#2-1.0}{#1};
}

\newcommand{\cornerIulptens}[2]{
\lineCul{#1-0.5}{#2+0.5};
\lineCul{#1+0.5}{#2-0.5};
\lineH{#1}{#1+1.5}{#2+0.5};
\lineH{#1+1.0}{#1+1.5}{#2-0.5};
\lineV{#2}{#2-1.5}{#1-0.5};
\lineV{#2-1.0}{#2-1.5}{#1+0.5};
}

\newcommand{\cornerIurptens}[2]{
\lineCur{#1+0.5}{#2+0.5};
\lineCur{#1-0.5}{#2-0.5};
\lineH{#1}{#1-1.5}{#2+0.5};
\lineH{#1-1.0}{#1-1.5}{#2-0.5};
\lineV{#2}{#2-1.5}{#1+0.5};
\lineV{#2-1.0}{#2-1.5}{#1-0.5};
}

\newcommand{\cornerIdlptens}[2]{
\lineCdl{#1+0.5}{#2+0.5};
\lineCdl{#1-0.5}{#2-0.5};
\lineH{#1}{#1+1.5}{#2-0.5};
\lineH{#1+1.0}{#1+1.5}{#2+0.5};
\lineV{#2}{#2+1.5}{#1-0.5};
\lineV{#2+1.0}{#2+1.5}{#1+0.5};
}

\newcommand{\cornerIdrptens}[2]{
\lineCdr{#1+0.5}{#2-0.5};
\lineCdr{#1-0.5}{#2+0.5};
\lineH{#1}{#1-1.5}{#2-0.5};
\lineH{#1-1.0}{#1-1.5}{#2+0.5};
\lineV{#2}{#2+1.5}{#1+0.5};
\lineV{#2+1.0}{#2+1.5}{#1-0.5};
}

\newcommand{\centerCudptens}[3]{
\cornerCp{#1}{#2}{#3};
\lineH{#1+1.0}{#1+1.5}{#2-0.5}
\lineH{#1+1.0}{#1+1.5}{#2+0.5}
\lineH{#1-1.0}{#1-1.5}{#2-0.5}
\lineH{#1-1.0}{#1-1.5}{#2+0.5}
}

\newcommand{\gaugeXultens}[3]{
\gaugeX{#1}{#2}{#3};
\lineH{#1+0.5}{#1+1.0}{#2};
\lineV{#2-0.5}{#2-1.0}{#1};
}

\newcommand{\cornerCurtens}[3]{
\cornerC{#1}{#2}{#3};
\lineH{#1-0.5}{#1-1.0}{#2};
\lineV{#2-0.5}{#2-1.0}{#1};
}

\newcommand{\cornerCdltens}[3]{
\cornerC{#1}{#2}{#3};
\lineH{#1+0.5}{#1+1.0}{#2};
\lineV{#2+0.5}{#2+1.0}{#1};
}

\newcommand{\cornerCdrtens}[3]{
\cornerC{#1}{#2}{#3};
\lineH{#1-0.5}{#1-1.0}{#2};
\lineV{#2+0.5}{#2+1.0}{#1};
}

\begin{document}
\title{Faster Methods for Contracting Infinite Two-Dimensional Tensor Networks}
\author{M.T. \surname{Fishman}}
\affiliation{Institute for Quantum Information and Matter, California Institute of Technology, 
Pasadena, California 91125, USA}
\author{L. \surname{Vanderstraeten}}
\affiliation{Ghent University, Faculty of Physics, Krijgslaan 281, 9000 Gent, Belgium}
\author{V. \surname{Zauner-Stauber}}
\affiliation{Vienna Center for Quantum Technology, University of Vienna, 
Boltzmanngasse 5, 1090 Wien, Austria}
\author{J. \surname{Haegeman}}
\affiliation{Ghent University, Faculty of Physics, Krijgslaan 281, 9000 Gent, Belgium}
\author{F. \surname{Verstraete}}
\affiliation{Vienna Center for Quantum Technology, University of Vienna, 
Boltzmanngasse 5, 1090 Wien, Austria}
\affiliation{Ghent University, Faculty of Physics, Krijgslaan 281, 9000 Gent, Belgium}
 
\begin{abstract}
We revisit the corner transfer matrix renormalization group (CTMRG) method of Nishino and Okunishi 
for contracting two-dimensional (2D) tensor networks and demonstrate that its performance can be 
substantially improved by determining the tensors using an eigenvalue solver as opposed to the 
power method used in CTMRG. 
We also generalize the variational uniform matrix product state (VUMPS) ansatz for 
diagonalizing 1D quantum Hamiltonians to the case of 2D transfer matrices and discuss similarities 
with the corner methods. 
These two new algorithms will be crucial to improving the performance of variational infinite 
projected entangled pair state (PEPS) methods. 
\end{abstract}
\maketitle

\section{Introduction}
\label{sec:intro}

Two-dimensional (2D) tensor networks are ubiquitous in many-body physics~\cite{Haegeman17}. 
They occur naturally in the context 2D classical many-body systems as representations of partition 
functions~\cite{Kramers41,Baxter68,Baxter78,Baxter82,Nishino95,Nishino96,Nishino97} and can 
represent ground states, finite temperature states and the time evolution of 1D quantum systems, 
e.g. for systems with local interactions in terms of Trotter-Suzuki decompositions~\cite{Trotter59,
Suzuki76,Suzuki85,Suzuki90,Wang97,Suzuki91,Bursill96,Vidal03,Vidal07}.
Additionally, they occur in the context of tensor product state (TPS)~\cite{Nishino00,Nishino01,
Gendiar03,Nishino04} or projected entangled pair state (PEPS)~\cite{Verstraete04} representations of 
2D quantum systems and boundaries of 3D classical systems.
Most 2D tensor networks of interest do not allow exact solutions and can only be studied 
approximately, and a copious array of numerical tensor network methods have been developed 
over many decades for their study~\cite{Kramers41,Baxter68,Baxter78,Baxter82,White92,White93,
Nishino95,Nishino95b,Nishino96,Bursill96,Nishino97,Wang97,Shibata97,Vidal03,Verstraete04,Levin07,Vidal07,
Orus08,McCulloch08,Xie09,Huang11a,Huang11b,Xie12,Evenbly15,Yang17,Bal17}.

Methods for contracting 2D tensor networks fall roughly into two main categories, which we 
refer to as ``coarse graining methods" and ``boundary methods."
Examples of coarse graining methods are tensor renormalization group (TRG)~\cite{Levin07} and 
extensions such as second renormalization group (SRG)~\cite{Xie09},
higher order tensor renormalization group (HOTRG)~\cite{Xie12}, and tensor network renormalization 
(TNR)~\cite{Evenbly15,Yang17,Bal17}.
A common feature of these methods is that the local degrees of freedom are combined and truncated, 
so the Hilbert space of the network is explicitly changed at each step.
For boundary methods, a matrix product state (MPS) is used as an ansatz for the environment, and this 
MPS is optimized in various ways.
Boundary methods include the density matrix renormalization group (DMRG) algorithm~\cite{White92,White93,
Nishino95,Bursill96,Wang97,Shibata97,McCulloch08}, the corner transfer matrix renormalization group 
(CTMRG) algorithm~\cite{Baxter68,Baxter78,Baxter82,Nishino96,Nishino97}, 
the time evolving block decimation 
(TEBD) algorithm~\cite{Vidal03,Vidal07,Orus08,Zaletel15}, the time dependent variational principle 
(TDVP)~\cite{Haegeman11,Haegeman16}, etc.
Boundary methods have certain advantages: they are optimized iteratively instead of optimized 
layer by layer like most coarse graining methods, the form of the environments can make it much 
easier to calculate arbitrary correlation functions, and they appear to be very well-suited for 
performing PEPS calculations~\cite{Verstraete04,Jordan08,Jordan09,Vanderstraeten15,Corboz16b,Vanderstraeten16}.

The history of modern boundary methods goes back to Nishino's application of DMRG to calculating
fixed points of transfer matrices~\cite{Nishino95}.
Soon after, Nishino and Okunishi created the CTMRG algorithm~\cite{Nishino96,Nishino97}
by combining the corner transfer matrix (CTM) method of Baxter~\cite{Baxter68,Baxter78,Baxter82} and 
White's DMRG algorithm~\cite{White92,White93}.
CTMRG was initially introduced as a powerful numerical tool for contracting 2D classical partition 
functions. 
In addition, it has been used extensively in TPS/PEPS calculations of 3D classical and 
2D quantum systems, where it is used to approximate the contraction of 2D tensor networks that
arise in those calculations.
CTMRG was used as the contraction method in the original TPS calculations~\cite{Nishino01,
Gendiar03,Maeshima01,Nishino04}. 
An MPS-based boundary method was used for the original finite PEPS calculation~\cite{Verstraete04}
while iTEBD, an MPS-based power method, was used to perform the original infinite PEPS~\cite{Jordan08,Jordan09} 
calculations.
Since then, PEPS calculations in the thermodynamic limit have mostly been performed using CTMRG as the 
contraction method, and a variety of advancements have been made to the method over recent years
in that context~\cite{Orus09,Corboz10a,Corboz10b,Corboz11,Orus12,Corboz14,Phien15,
Corboz16a,Corboz16b}.

Here, we present two new approaches that improve upon the speed of CTMRG for contracting 2D tensor 
networks in the thermodynamic limit.
First, we present a transfer matrix version of the recently introduced variational uniform matrix 
product state (VUMPS)~\cite{ZaunerStauber18} algorithm for contracting 2D tensor networks.
We also present a new corner method analogous to CTMRG that better exploits translational 
invariance by solving for the environment tensors using a set of fixed point equations.
We present benchmark results for VUMPS and our new corner method, showing remarkable speedups over 
CTMRG, particularly for systems near criticality.
Our benchmarks include a variety of both 2D statistical mechanics models and 2D quantum systems 
represented as PEPS.

\section{Problem Statement}

We are interested in the approximate numerical contraction of infinite 2D tensor networks. 
For simplicity, throughout the paper we will focus on tensor networks on an infinite square 
lattice with a single site unit cell.
We are agnostic about where the tensor network comes from: it could be a 2D classical partition 
function, the norm of a PEPS, etc. 

For concreteness, we are interested in evaluating the contraction of the following tensor 
network
\begin{equation}
  \begin{tikzpicture}[every node/.style={scale=0.9},scale=.45]
  \tracetwodimnetwork{0.0}{0.0}
  \end{tikzpicture}
\label{eq:partfnc}
\end{equation}
(for readers unfamiliar with tensor networks, we refer them to Ref.~\onlinecite{Orus14} for an 
introduction).
In Eq.~\eqref{eq:partfnc}, we work directly in the thermodynamic limit, i.e. the number of 
lattice sites in the horizontal and vertical directions, $M,N$, approaches $\infty$. 
$Tr[...]$ denotes two traces, one over the open horizontal indices and another over the open 
vertical indices.
If the network represents a 2D classical partition function, the fourth-order tensor $T$ is related 
to the local Boltzmann weight (possibly up to a local tensor renormalization) and $\kappa$ is defined
to be the ``partition function per site,"~\cite{Baxter82} related to the free energy per site. 
If the network is the evaluation of the norm of a PEPS, each tensor $T$ is the bra and ket
PEPS tensor at each site contracted over the physical index\footnote{The PEPS tensors can of course
be left uncontracted to allow for a more efficient ordering of contraction later on, but for now we 
will think of it as a single larger tensor.}, and $\kappa$ is the norm per site.

We are also interested in calculating observables such as expectation values of local operators or 
correlation functions. 
In terms of the tensor network, these are represented as impurity sites, such as:
\begin{equation}
  \begin{tikzpicture}[every node/.style={scale=0.90},scale=.45]
  \tracetwodimnetworkXY{0.0}{0.0}
  \end{tikzpicture}.
  \label{eq:expval}
\end{equation}
We want a contraction method that makes it easy to calculate arbitrary correlation functions, 
since they show up in e.g. calculating structure factors or summing Hamiltonian terms in 
variational PEPS ground state optimizations\cite{Vanderstraeten15,Corboz16b,Vanderstraeten16}.
For this reason we focus on MPS boundary methods, which make it much easier to calculate arbitrary 
correlation functions.
It is more challenging in methods like TRG/TNR where all of the tensors at each layer must 
properly be kept track of, and calculating arbitrary correlation functions on the lattice is
potentially very complicated.

Here we will also define the row-to-row transfer matrix, which is simply a single infinite row
of the tensor network:
\begin{equation}
  \begin{tikzpicture}[every node/.style={scale=0.90},scale=.45]
  \rowtorow{0.0}{0.0}
  \end{tikzpicture}.
  \label{eq:rowtorow}
\end{equation}
The row-to-row transfer matrix is an infinite, translationally invariant matrix product operator
(MPO).
We also define the column-to-column transfer matrix as an infinite column of the tensor
network:
\begin{equation}
  \begin{tikzpicture}[every node/.style={scale=0.90},scale=.45]
  \coltocol{0.0}{0.0}
  \end{tikzpicture}.
  \label{eq:coltocol}
\end{equation}

For MPS boundary methods, the evaluation of diagrams like Eq.~\eqref{eq:partfnc}--\eqref{eq:expval} 
is performed by finding the leading up and down eigenvectors of the row-to-row transfer matrix
portrayed in \eqref{eq:rowtorow} and the leading left and right eigenvectors of column-to-column 
transfer matrix portrayed in \eqref{eq:coltocol}.
Exact MPS representations of these eigenvectors are in general infinitely large, but for many 
2D tensor networks representing physical many-body systems good MPS approximations exist (in some 
cases provably\cite{Hastings07}).
We refer to these uniform MPS fixed points as the up, down, left and right boundary MPSs, 
and call their MPS tensors respectively $A_U$, $A_D$, $A_L$ and $A_R$.
As an example, the fixed point equation for the top MPS is as follows:
\begin{equation}
  \begin{tikzpicture}[every node/.style={scale=0.90},scale=.45]
  \mpsUpFour{0.0}{2.0}
  \rowtorow{0.0}{0.0}
  \draw (9.0,1.0) node (X) {$\approx$};
  \draw (0.0,-2.0) node (X) {$\kappa^M$};
  \mpsUpFour{4.0}{-2.0}
  \end{tikzpicture}
  \label{eq:topfixedpoint}
\end{equation}
where $\kappa$ is the partition function or norm per site defined in Eq.~\eqref{eq:partfnc}.
At the fixed point, analogous equations to Eq.~\eqref{eq:topfixedpoint} should be satisfied
by the other boundary MPSs.

Once the boundary MPSs are obtained, local observables and correlations functions can be
computed efficiently.
For example, the expectation value of a local observable can be calculated from the
up and down boundary MPSs as follows:
\begin{equation}
  \begin{tikzpicture}[every node/.style={scale=0.90},scale=.45]
  \draw[scale=1.0] (-4.5,0.0) node (X) {$\langle X\rangle = $};
  \mpsUpFive{0.0}{5.0}
  \rowtorowFiveX{0.0}{3.0}
  \mpsDownFive{0.0}{1.0}
  \lineH{-3.0}{11.0}{0.0}
  \mpsUpFive{0.0}{-1.0}
  \rowtorowFive{0.0}{-3.0}
  \mpsDownFive{0.0}{-5.0}
  \end{tikzpicture}
  \label{eq:localexpvalue}.
\end{equation}
These networks can then be contracted efficiently\cite{Orus08}.
Arbitrary correlation functions can also be computed efficiently using fixed point
boundary MPSs\cite{Vanderstraeten15}.

There are many different approaches to obtaining the four boundary MPS fixed points of the
row-to-row transfer matrix \eqref{eq:rowtorow} and column-to-column transfer matrix 
\eqref{eq:coltocol}.
In the next section, we review one very commonly used contraction method, the corner transfer 
renormalization group (CTMRG) algorithm of Nishino and Okunishi\cite{Baxter68,Baxter78,Baxter82,
Nishino96,Nishino97}, and describe two new proposals, one based on the recently proposed variational 
uniform matrix product state (VUMPS) algorithm\cite{ZaunerStauber18}, and one that we refer
to as the fixed point corner method (FPCM), which is like CTMRG but solves for the boundary
tensors using a series of fixed point equations.

\section{Algorithm overview}
\label{sec:algorithmoverview}

One strategy for evaluating Eq.~\eqref{eq:partfnc}--\eqref{eq:expval} involves finding a single
boundary MPS eigenvector at a time. 
The infinite time evolving block decimation (iTEBD) of Or\'us and Vidal~\cite{Vidal03,Vidal07,Orus08}
is an example of this strategy. 
In iTEBD, a power method is used to find the fixed point MPS eigenvector by repeatedly applying a 
row-to-row or column-to-column transfer matrix to a starting MPS.
In this work, one of the strategies we propose also focuses on solving for a single MPS eigenvector
for each direction at a time.
Instead of iTEBD, we propose using the recently introduced variational uniform matrix product state 
(VUMPS) algorithm~\cite{ZaunerStauber18}, which can be viewed as an improvement on the infinite 
density matrix renormalization group (iDMRG)~\cite{White92,White93,McCulloch08} where an MPS
is optimized directly in the thermodynamic limit instead of grown site-by-site to reach the
thermodynamic limit.
VUMPS was originally applied to finding ground state approximations of 1D and quasi-1D quantum states,
where it was shown to substantially improve the computational performance over iTEBD
and iDMRG.
In analogy to how Nishino introduced the transfer matrix DMRG (TMRG) method as an extension of 
applying DMRG to finding the fixed point MPS approximation of the transfer matrices of partition
functions~\cite{Nishino95}, in Section~\ref{subsec:vumps} we show how VUMPS can be applied to find 
the fixed point of infinite uniform transfer matrices.

Another strategy for finding the boundary MPSs is to attempt to find all four MPSs at once.
An example of this approach is the corner transfer matrix (CTM) method of Baxter~\cite{Baxter68,
Baxter78,Baxter82}, and its improvement by Nishino and Okunishi called the corner transfer matrix
renormalization group (CTMRG)~\cite{Nishino96,Nishino97}.
In the CTMRG algorithm, all four boundary MPSs are iteratively optimized.
We give a brief review of CTMRG in Section~\ref{subsec:ctmrg}.
One of the new methods we propose in this work, which we refer to as the fixed point corner method 
(FPCM) and is explained in Section~\ref{subsec:fpcm}, also solves for all four MPS fixed points at 
once.
Like CTMRG, FPCM uses CTMs, but solves for the CTMs and MPS tensors using a series of fixed
point equations (which were originally written down by Baxter~\cite{Baxter68,Baxter78,Baxter82}).

\subsection{VUMPS for contracting infinite 2D tensor networks}
\label{subsec:vumps}

Here, we present the application of the recently proposed VUMPS algorithm~\cite{ZaunerStauber18} 
to finding MPS fixed points of infinitely large, translationally invariant transfer matrices.
Essentially, we apply VUMPS to the problem of directly finding fixed points of the form
shown in Eq.~\eqref{eq:topfixedpoint}.

We now present VUMPS for obtaining the top fixed point MPS of the network.
We would like to find the uniform MPS satisfying Eq.~\eqref{eq:topfixedpoint}.
In VUMPS, we use the mixed canonical form of the MPS, so Eq.~\eqref{eq:topfixedpoint} becomes:
\begin{equation}
  \begin{tikzpicture}[every node/.style={scale=0.90},scale=.55]
  \mpsFiveMixed{0.0}{2.0}
  \draw (-2.0,1.0) node (X) {$\dots$};
  \mpoOneFive{0.0}{0.0}
  \draw (10.0,1.0) node (X) {$\dots$};
  \draw (11.0,1.0) node (X) {$\propto$};
  \draw (-2.0,-3.0) node (X) {$\dots$};
  \mpsFiveMixed{0.0}{-3.0}
  \draw (10.0,-3.0) node (X) {$\dots$};
  \end{tikzpicture}.
  \label{eq:vumps}
\end{equation}
In the mixed canonical gauge, for the state to be (approximately) translationally invariant the tensors 
must satisfy the relations:
\begin{equation}
  \begin{tikzpicture}[every node/.style={scale=0.90},scale=.55]
   \MPSuLtens{0.0}{0.0}{A_U^L}
   \centerCudtens{2.0}{0.0}{C_U}
    \draw (4.0,0.0) node {$\approx$};
   \MPSuCtens{6.0}{0.0}{A_U^C}
    \draw (8.0,0.0) node {$\approx$};
   \centerCudtens{10.0}{0.0}{C_U}
   \MPSuRtens{12.0}{0.0}{A_U^R}
  \end{tikzpicture}
  \label{eq:vumps_pullthrough}
\end{equation}
where the singular values of the matrix $C_U$ are the Schmidt values of the uniform MPS.
Note the inequalities in Eq.~\eqref{eq:vumps_pullthrough}, since the relationships will 
not generally all simultaneously be satisfied exactly during the optimization.
How accurately they are satisfied will relate to how translationally invariant the state is,
and should be satisfied to very high accuracy at the fixed point of the VUMPS algorithm.
Additionally, $A_U^L$ and $A_U^R$ are isometric tensors satisfying:
\begin{equation}
  \begin{tikzpicture}[every node/.style={scale=1.0},scale=.65]
  \defLvumps{0.0}{0.0}{A_U^L}{\bar{A}_U^L}
  \end{tikzpicture}
\end{equation}
\begin{equation}
  \begin{tikzpicture}[every node/.style={scale=1.0},scale=.65]
  \defRvumps{6.0}{0.0}{A_U^R}{\bar{A}_U^R}
  \end{tikzpicture}
\end{equation}
at all times.
Any uniform MPS can be turned into this form, for example with the algorithm introduced in 
Ref.~\onlinecite{Orus08} in the context of iTEBD or with the algorithm introduced in 
Ref.~\onlinecite{Haegeman17} and expanded on in Appendix~\ref{sec:MPSgauging_sym}.

The VUMPS algorithm proceeds by repeating the following steps until convergence:
\begin{enumerate}
\item Solve for the environments:
\begin{equation}
  \begin{tikzpicture}[every node/.style={scale=1.0},scale=.65]
      \cornerIultens{0.0}{2.0}
      \MPSltens{0.0}{0.0}{E_L}
      \cornerIdltens{0.0}{-2.0}
      \MPOtens{2.0}{0.0}{T}
      \MPSuLtens{2.0}{2.0}{A_U^L}
      \MPSdLtens{2.0}{-2.0}{\bar{A}_U^L}
      \draw (4.0,0.0) node (X) {$\approx$};
      \draw (5.0,0.0) node (X) {$\kappa_L$};
      \cornerIultens{6.0}{2.0}
      \MPSltens{6.0}{0.0}{E_L}
      \cornerIdltens{6.0}{-2.0}
  \end{tikzpicture}
\end{equation}
\begin{equation}
  \begin{tikzpicture}[every node/.style={scale=1.0},scale=.65]
      \MPOtens{10.0}{0.0}{T}
      \MPSuRtens{10.0}{2.0}{A_U^R}
      \MPSdRtens{10.0}{-2.0}{\bar{A}_U^R}
      \cornerIurtens{12.0}{2.0}
      \MPSrtens{12.0}{0.0}{E_R}
      \cornerIdrtens{12.0}{-2.0}
      \draw (13.0,0.0) node (X) {$\approx$};
      \draw (14.0,0.0) node (X) {$\kappa_R$};
      \cornerIurtens{16.0}{2.0}
      \MPSrtens{16.0}{0.0}{E_R}
      \cornerIdrtens{16.0}{-2.0}
  \end{tikzpicture}
\end{equation}
where $\kappa_L\approx \kappa_R$ up to errors in Eq.~\eqref{eq:vumps_pullthrough}.
\label{alg:vumps_environment}
\item Solve for zero-site and single-site tensors:
\begin{equation}
  \begin{tikzpicture}[every node/.style={scale=0.90},scale=.55]
    \cornerIultens{0.0}{0.0}
    \centerCudtens{2.0}{0.0}{C_U}
    \cornerIurtens{4.0}{0.0}
    \MPSltens{0.0}{-2.0}{E_L}
    \centerIudtens{2.0}{-2.0}
    \MPSrtens{4.0}{-2.0}{E_R}
    \cornerIdrtens{0.0}{-4.0}
    \cornerIdltens{4.0}{-4.0}
    \draw (6.0,-2.0) node (X) {$\approx$};
    \draw (7.0,-2.0) node (X) {$\lambda_C$};
    \centerCudtens{9.0}{-2.0}{C_U}
 \end{tikzpicture}
\end{equation}
\begin{equation}
  \begin{tikzpicture}[every node/.style={scale=0.90},scale=.55]
    \cornerIultens{0.0}{0.0}
    \MPSuCtens{2.0}{0.0}{A_U^C}
    \cornerIurtens{4.0}{0.0}
    \MPSltens{0.0}{-2.0}{E_L}
    \MPOtens{2.0}{-2.0}{T}
    \MPSrtens{4.0}{-2.0}{E_R}
    \cornerIdrtens{0.0}{-4.0}
    \cornerIdltens{4.0}{-4.0}
    \Iudtens{2.0}{-4.0}
    \draw (6.0,-2.0) node (X) {$\approx$};
    \draw (7.0,-2.0) node (X) {$\lambda_{A^C}$};
    \MPSuCtens{9.0}{-2.0}{A_U^C}
  \end{tikzpicture}
\end{equation}
where $\lambda_{A^C}/\lambda_C \approx \kappa_{L/R}$ near or at the fixed point.
\label{alg:vumps_solveUC}
\item From $A_U^C$ and $C_U$ found in step ~\ref{alg:vumps_solveUC}, find new MPS tensors 
$A_U^L$ and $A_U^R$ satisfying Eq.~\eqref{eq:vumps_pullthrough}.
Techniques for numerically solving these equations are described in the original VUMPS proposal in 
Ref.~\onlinecite{ZaunerStauber18}.
\label{alg:vumps_solveULR}
\end{enumerate}

The VUMPS algorithm proceeds by repeating steps~\ref{alg:vumps_environment}--
\ref{alg:vumps_solveULR} until convergence. 
Convergence can be measured, for example, by the change in the singular
values of $C$ from step to step. 
Another measure for the convergence that can be used is the norm of the residual $B_U$:
\begin{equation}
  \begin{tikzpicture}[every node/.style={scale=0.75},scale=.4]
    \MPSuCtens{-4.0}{-4.0}{B_U}
    \draw (-2.0,-4.0) node (X) {$=$};
    \residualLHS{0.0}{0.0}
    \draw (6.0,-4.0) node (X) {$-$};
    \draw (7.0,-4.0) node (X) {$\kappa_L$};
    \residualRHS{11.0}{0.0}
 \end{tikzpicture}
\end{equation}
(or the analogous right version), similar to the gradient discussed in 
Ref.~\onlinecite{ZaunerStauber18}.

For finding the fixed point of a row-to-row or column-to-column transfer matrix Hermitian about the 
horizontal, this scheme maps directly to the original VUMPS proposal\cite{ZaunerStauber18},
and the algorithm solves for both the top and bottom fixed points, which are just Hermitian
conjugates of each other.
For a Hermitian row-to-row transfer matrix, the fixed points environments $E_L,E_R$ are related to the 
fixed points of the boundary MPS tensors $A_L,A_R$ used in the CTMRG ansatz (which we review in 
Section~\ref{subsec:ctmrg}), the gauged MPS 
tensors $A_U^L,A_U^R$ are related to the fixed points of isometric projectors used to renormalize 
the CTM environment (i.e. the eigenvectors of the product of the four CTMs), and the center tensor 
$C_U$ is related to the product of CTMs $C_{LU} C_{UR}$.
This correspondence is discussed in more detail in Ref.~\onlinecite{Haegeman17}.
A similar correspondence between the fixed point of CTMRG and the fixed point of DMRG applied
to Hermitian transfer matrices was pointed out by Nishino and Okunishi~\cite{Nishino95,Nishino96,Nishino97}.

For contracting 2D statistical mechanics partition functions and calculating the norm of a PEPS,
transfer matrix VUMPS is in fact simpler than the original proposal, because we do not in general 
have to be concerned about summing Hamiltonian terms which can lead to divergences if the fixed 
point is not calculated properly~\cite{Michel10,ZaunerStauber18}, and methods such as Arnoldi can 
be directly employed to find the fixed points.
One may have to be more careful contracting networks that involve sums of local 
operators, such as when calculating structure factors or gradients of PEPS. 
See Ref.~\onlinecite{Corboz16b,Vanderstraeten15,Vanderstraeten16} for approaches to contracting 
such networks, where the environments calculated from the norm of a PEPS are used to aid in the 
contraction.

In general for a non-Hermitian network, to get the environment for calculating local observables, one must 
additionally solve for the bottom fixed point (and in order to calculate arbitrary correlation 
functions, the left and right fixed point MPSs as well). 
For a network that isn't ``very asymmetric," the top fixed point can be used as a good starting 
point for the bottom fixed point MPS. 
It is also important to note that in the case of symmetry breaking, one should take care that the
fixed points in different directions are all compatible, in the sense that they correspond to the 
same symmetry-broken states.

For non-Hermitian networks, the method we propose here is analagous to iTEBD, where each of the four
boundary MPSs is solved for in separate optimizations (although in iTEBD
the fixed points in each direction are obtained with power methods, which was shown to be
slower than VUMPS in Ref.~\onlinecite{ZaunerStauber18}).
An alternative approach from the one proposed here is to solve for two opposing fixed points in the 
same optimization (for example both the top and bottom fixed points of the row-to-row transfer matrix).
This is approach has been used in the context of applying DMRG to non-Hermitian transfer
matrices (TMRG)~\cite{Bursill96,Wang97,Shibata97,Kemper01,Enss01,Chan05,Schollwock05,Huang11a,Huang11b,Huang12}.
It would be interesting to generalize the transfer matrix VUMPS algorithm to solving for both fixed 
points at once, but we do not explore that here.

\subsection{Corner transfer matrix renormalization group (CTMRG) review}
\label{subsec:ctmrg}

In this section we review the corner transfer matrix renormalization group algorithm, which
was originally introduced by Nishino and Okunishi and extended in a variety of other works
in the context of PEPS calculations.
The general ansatz used for the environment in the corner transfer matrix renormalization group 
(CTMRG) algorithm is as follows:
\begin{equation}
  \begin{tikzpicture}[every node/.style={scale=0.8},scale=.55]
  \ctmrgansatzTwoTwo{0.0}{0.0}
  \end{tikzpicture}.
  \label{eq:ctmansatz}
\end{equation}
The matrices $\{C_i\}$ in Eq.~\eqref{eq:ctmansatz}, known as the corner transfer 
matrices (CTMs), were originally introduced by Baxter for studying 2D classical
statistical mechanics problems\cite{Baxter68,Baxter78,Baxter82,Nishino96}.
The CTMs represent approximations of the infinite corners of the tensor network.
The boundary MPS tensors $\{A_i\}$ in Eq.~\eqref{eq:ctmansatz} represent approximations of the 
half-row transfer matrices (HRTMs) and half-column transfer matrices (HCTMs).
In our notation, $C_{LU}$ denotes the CTM approximating the upper left corner of the 
network, $A_L$ denotes the left HRTM of the network, $A_U$ denotes the upper HCTM of the
network, etc.
We refer to the set of tensors $\{C_i,A_j\}$ as the environment of the 2D tensor network.

The CTMRG algorithm is thought of in terms of contracting row-to-row transfer matrices 
and/or column-to-column transfer matrices composed of tensor $T$ into the environment, 
either simultaneously in multiple directions or sequentially in specified
directions (depending on details of the renormalization scheme).

If the row-to-row and column-to-column transfer matrices of the network are absorbed into 
the environment indefinitely, then the environment tensors would grow exponentially in size, 
so some sort of truncation scheme is required. 
The truncation is referred to as renormalization.
This renormalization of the enlarged environment is performed by introducing projectors into the 
network.
There are multiple methods available for how to grow the lattice as well as how to choose the 
projectors.
We will start by describing how these projectors are chosen for tensor networks with reflection
symmetries, where the ansatz in Eq.~\eqref{eq:ctmansatz} can be constrained.

\subsubsection{Symmetric CTMRG review}
\label{subsubsec:ctmrg_sym}

To get some intuition for how CTMRG works, it is useful to discuss the case in which the network
tensor $T$ is Hermitian about all reflections (about the horizontal, vertical and 
diagonals, in other words $T_{lurd}=\bar{T}_{ruld}=\bar{T}_{ldru}=\bar{T}_{drul}$\footnote{Note
that this is equivalent to the MPO tensor being invariant under $\pi/2$ rotations.}).
This is the case for many statistical mechanics models.
In this case, we can constrain the environment tensors in the ansatz in Eq.~\eqref{eq:ctmansatz}
to satisfy $A_U=A_R=A_D=A_L\equiv A$, $C_{LU}=C_{UR}=C_{RD}=C_{DL}\equiv C$, and 
additionally impose $C=C^{\dagger}$ and $A^s=(A^s)^{\dagger}$\footnote{Note that we use the notation
$A^s$ to denote the matrix obtained by setting the physical index of MPS tensor $A$ to $s$.}.
Eq.~\eqref{eq:ctmansatz} becomes:
\begin{equation}
  \begin{tikzpicture}[every node/.style={scale=0.9},scale=.55]
  \ctmrgansatzTwoTwoSym{0.0}{0.0}
  \end{tikzpicture}.
  \label{eq:ctmansatzsym}
\end{equation}

This is the CTMRG case that was covered in the initial proposal of Nishino and 
Okunishi~\cite{Nishino96,Nishino97} (though extensions to the asymmetric case were discussed).
The CTMRG algorithm consists of obtaining the projector by ``growing" the corner transfer matrices 
$C$ by absorbing surrounding network and environment tensors and performing a Hermitian 
eigendecomposition, and we summarize the algorithm here:
\begin{enumerate}
\item Obtain the projector from a Hermitian eigendecomposition of the grown corner transfer matrix
\footnote{The original proposal actually involved a symmetric eigendecomposition of a product of 
four of the grown corners in Eq.~\eqref{eq:ctmsym_eigdecomp} which has the interpretation of
a density matrix, but the eigenbasis is the same as that of a single corner.}:
\begin{equation}
  \begin{tikzpicture}[every node/.style={scale=0.9},scale=.55]
  \CULTsym{0.0}{0.0}
  \draw[scale=1.0] (2.0,0.0) node (X) {$\approx$};
  \CULsymeig{7.0}{0.0}
  \end{tikzpicture}
  \label{eq:ctmsym_eigdecomp}
\end{equation}
where we use the convention that the indices of the tensor in the diagram are ordered clockwise,
except when the complex conjugate is taken in which case the ordering is reversed.
In Eq.~\eqref{eq:ctmsym_eigdecomp}, the tensor network on the left side is contracted, reshaped 
into a Hermitian matrix, a Hermitian eigendecomposition is performed, and the bond dimension 
is truncated according to the eigenvalues.
$D$ is a diagonal matrix storing the largest magnitude (real) eigenvalues.
$U$ is the matrix of the orthonormal eigenvectors associated with the largest eigenvalues $D$
reshaped into an isometric tensor. 
$U$ satisfies $(U^s)^{\dagger} U^s = I$ (using Einstein summation convention) or diagramatically:
\begin{equation}
  \begin{tikzpicture}[every node/.style={scale=0.95},scale=.55]
  \defPLsym{0.0}{0.0}
  \end{tikzpicture}.
  \label{eq:defPLsym}
\end{equation}

\label{alg:ctmsym_eigdecomp}
\item Renormalize the grown environment.
The new CTMRG environment is obtained by absorbing a row and column of the tensor network in 
each direction into the environment. 
The renormalization is performed with the projector $U^s(U^{s'})^{\dagger}$, which diagramatically
is:
\begin{equation}
  \begin{tikzpicture}[every node/.style={scale=0.95},scale=.55]
  \defprojLsym{0.0}{0.0}
  \end{tikzpicture}.
  \label{eq:defprojLsym}
\end{equation}
The projector Eq.~\eqref{eq:defprojLsym} is inserted into the grown boundary environment at 
every link in the environment, and grown environment tensors are renormalized to obtain the new 
environment tensors.
The new environment tensors $C'$ and $A'$ are obtained as follows:
\begin{equation}
  \begin{tikzpicture}[every node/.style={scale=0.9},scale=.55]
  \cornerCultens{0.0}{0.0}{C'}
  \draw[scale=1.0] (2.0,0.0) node (X) {$=$};
  \CULTrenormsym{6.0}{0.0}
  \end{tikzpicture}
  \label{eq:ctmsym_CLUrenorm}
\end{equation}
\begin{equation}
  \begin{tikzpicture}[every node/.style={scale=0.95},scale=.55]
    \ALpsym{0.0}{0.0}
    \draw (2.0,0.0) node (X) {$=$};
    \ALrenormsym{6.0}{0.0}
  \end{tikzpicture}.
  \label{eq:ctmsym_ALrenorm}
\end{equation}
Of course, from Eq.~\eqref{eq:ctmsym_eigdecomp}, we can trivially see that $C'=D$, but the
more general form of Eq.~\eqref{eq:ctmsym_CLUrenorm} will be useful when we discuss generalizing
to situations where the tensor network is comprised of asymmetric tensors $T$, and when we
discuss our new fixed point corner method.
\label{alg:ctmsym_renorm}
\end{enumerate}

The CTMRG algorithm essentially involves iterating steps~\ref{alg:ctmsym_eigdecomp} 
and ~\ref{alg:ctmsym_renorm} until convergence (for example, measured by the difference in 
the eigenvalues of the corner transfer matrices between steps), where one must make sure to
normalize the HRTMs, HCTMs, and CTMs at each steps.

Extensions to networks with other types of symmetries are straightforward.
If the network is Hermitian about the horizontal and vertical directions but not the diagonal
directions, we can impose $A_U=A_D\equiv A$ (where $A=A^{\dagger}$) and $A_L=A_R\equiv B$ (where 
$B=B^{\dagger}$), and $C_{LU}=C_{UR}^{\dagger}=C_{RD}=C_{DL}^{\dagger}\equiv C$ (where 
$C=C^{\dagger}$ if the gauge is chosen properly).
In that case, a generalization of Eq.~\eqref{eq:ctmsym_eigdecomp} can be used to obtain 
projectors for the left/right direction and up/down direction using the left and right singular vectors 
respectively obtained from the SVD of one of the grown corners.

If the tensor network is not Hermitian about either a horizontal or vertical reflection, 
a simple Hermitian eigendecomposition like that shown in Eq.~\eqref{eq:ctmsym_eigdecomp}
will not suffice, and a more involved scheme must be invoked.
In general, the projectors used will not be isometric, and there will be different projectors for 
renormalizing each direction (left, up, right, and down).
We refer readers to Ref.~\onlinecite{Orus12} for a discussion of applying CTMRG to the
case when the network is Hermitian about a single direction.
In the next section, we will discuss strategies for generalizing CTMRG to fully asymmetric tensor 
networks.

\subsubsection{Asymmetric CTMRG review}
\label{subsubsec:ctmrg_nonsym}

When the tensor network does not contain reflection symmetries as discussed in the previous
section, in general there is not a unique way for choosing the projector and a variety of methods
have been proposed.
In the previous section, we discussed methods where the network was renormalized in all four 
directions at once.
Here, we will focus on what is called the ``directional" approach, where a single direction
of the network is renormalized at a time, and a single CTMRG step constitutes cycling
through the different directions.
This approach makes the discussion easier, and is well-suited for PEPS calculations~\cite{Corboz11,
Corboz14}.

In the directional approach, the ``left move" involves contracting just the column-to-column transfer 
matrix into the left environment, and renormalizing with a projector which we will call
$P_L^s [P_L^-]^{s'}$ or diagramatically:
\begin{equation}
  \begin{tikzpicture}[every node/.style={scale=0.95},scale=.55]
  \defprojL{0.0}{0.0}
  \end{tikzpicture}.
  \label{eq:defprojL}
\end{equation}
$P_L^-$ denotes the approximate left inverse of $P_L$, in other words they satisfy 
$[P_L^-]^s P_L^s \approx I$ or diagramatically:
\begin{equation}
  \begin{tikzpicture}[every node/.style={scale=0.95},scale=.55]
  \defPL{0.0}{0.0}
  \end{tikzpicture}.
  \label{eq:defPL}
\end{equation}

Using the projector in Eq.~\eqref{eq:defprojL}, the left move is shown below:
\begin{equation}
  \begin{tikzpicture}[every node/.style={scale=0.85},scale=.55]
  \cornerCultens{0.0}{0.0}{C_{LU}'}
  \draw[scale=1.0] (2.0,0.0) node (X) {$=$};
  \CULAUrenorm{6.0}{1.0}
  \end{tikzpicture}
  \label{eq:ctm_CLUrenorm}
\end{equation}
\begin{equation}
  \begin{tikzpicture}[every node/.style={scale=0.95},scale=.55]
    \ALp{0.0}{0.0}
    \draw (2.0,0.0) node (X) {$=$};
    \ALrenorm{6.0}{0.0}
  \end{tikzpicture}
  \label{eq:ctm_ALrenorm}
\end{equation}
\begin{equation}
  \begin{tikzpicture}[every node/.style={scale=0.85},scale=.55]
  \cornerCdltens{0.0}{0.0}{C_{DL}'}
  \draw[scale=1.0] (2.0,0.0) node (X) {$=$};
  \CDLADrenorm{6.0}{-1.0}
  \end{tikzpicture}.
  \label{eq:ctm_CLDrenorm}
\end{equation}
The diagrammatic notation we use for the tensors in the projector is suggestive, identifying them
as MPS tensors. 
This notation will prove useful later on when we present our new algorithms.
The up, right and down moves are simply rotated versions of the left move.

The projectors (i.e. Eq.~\eqref{eq:defprojL}) are obtained from the current guess for the environment, 
and in general the choice is not unique.
Many methods for obtaining these projectors have been proposed over the 
years~\cite{Nishino96,Orus09,Corboz10a,Corboz10b,Corboz11,Orus12,Corboz14}.
In this work, for asymmetric CTMRG, we will use the method that is most commonly used in modern 
infinite PEPS calculations, the one proposed in Ref.~\onlinecite{Corboz14}.
We summarize the left move of that asymmetric CTMRG method here:
\begin{enumerate}
\item The first step of the left move of the CTMRG method proposed in Ref.~\onlinecite{Corboz14} 
  is to perform the following QR decompositions:
\begin{equation}
  \begin{tikzpicture}[every node/.style={scale=0.7},scale=.48]
  \CULT{0.0}{0.0}
  \CURT{2.0}{0.0}
  \CDLT{0.0}{-3.0}
  \CDRT{2.0}{-3.0}
  \draw[scale=1.0] (5.5,-2.0) node (X) {$\approx$};
  \QRU{10.0}{-0.5}
  \QRD{10.0}{-3.5}
  \end{tikzpicture}
  \label{eq:QRfactorize}
\end{equation}
\item Next we obtain the tensors $P_L,P_L^-$ from $R_U$ and $R_D$ obtained in
  previous step.
  This is done using a method we will refer to as ``biorthogonalization" which was originally 
  proposed in the context of transfer matrix DMRG (TMRG) in Ref.~\onlinecite{Huang11a,Huang11b}.
  The first step of the biorthogonalization procedure is to perform the following SVD:
\begin{equation}
  \begin{tikzpicture}[every node/.style={scale=0.85},scale=.55]
  \cornerIultens{0.0}{0.0}
  \cornerIdltens{0.0}{-2.0}
  \MPStransferC{2.0}{0.0}{R_U^T}{R_D}
  \draw[scale=1.0] (4.0,-1.0) node (X) {$\approx$};
  \cornerCultens{6.0}{0.0}{\Sigma_L^2}
  \cornerIdltens{6.0}{-2.0}
  \gaugeARudtens{8.0}{0.0}{\bar{Q}_L}
  \gaugeARudtens{8.0}{-2.0}{W_L}
  \end{tikzpicture}
  \label{eq:biortho_corboz}
\end{equation}
where the second line is obtained by taking the SVD
$R_D^s [R_U^s]^T = W_L \Sigma_L^2 Q_L^{\dagger}$ and truncating according to the singular
values to the desired bond dimension.
\item Finally, we obtain $P_L,P_L^-$:
\begin{equation}
  \begin{tikzpicture}[every node/.style={scale=0.9},scale=.55]
  \MPSuLtens{0.0}{0.0}{P_L}
  \draw[scale=1.0] (2.0,0.0) node (X) {$=$};
  \MPSuCtens{4.0}{0.0}{R_U^T}
  \gaugeALudtens{6.0}{0.0}{Q_L}
  \centerCudtens{8.0}{0.0}{\Sigma_L^+}
  \end{tikzpicture}
  \label{eq:biortho_corboz_PL}
\end{equation}
\begin{equation}
  \begin{tikzpicture}[every node/.style={scale=0.9},scale=.55]
  \MPSdLtens{0.0}{0.0}{P_L^-}
  \draw[scale=1.0] (2.0,0.0) node (X) {$=$};
  \MPSdCtens{4.0}{0.0}{R_D}
  \gaugeALudtens{6.0}{0.0}{\bar{W}_L}
  \centerCudtens{8.0}{0.0}{\Sigma_L^+}
  \end{tikzpicture}
  \label{eq:biortho_corboz_tildePL}
\end{equation}
Note that it may be necessary for stability to use a pseudoinverse of $\Sigma_L$, where we denote
the pseudoinverse with $\Sigma_L^+$.
\end{enumerate}

In Appendix~\ref{sec:CTMRG_asym_alts} we discuss technical details about the stability of this method
as well as some new alternatives.
In the next section, we discuss a transfer matrix version of the variational uniform matrix product
state (VUMPS) algorithm introduced in Ref.~\onlinecite{ZaunerStauber18}.

\subsection{New fixed point corner method (FPCM)}
\label{subsec:fpcm}

Here we present a new corner method, which we refer to as the fixed point corner method (FPCM), 
which is similar to the CTMRG algorithm but solves for environment tensors in terms of fixed points.

\subsubsection{Symmetric FPCM}
\label{subsubsec:fpcm_sym}

We start with the simplest version of our new fixed point corner method (FPCM), when the
network is comprised of a tensor $T$ that is Hermitian about all reflections.
In this case, we use the same ansatz for the environment as we would use for the fully symmetric 
CTMRG algorithm, which we mentioned previously in Section~\ref{subsubsec:ctmrg_sym} and repeated here:
\begin{equation}
  \begin{tikzpicture}[every node/.style={scale=0.90},scale=.55]
  \ctmrgansatzTwoTwoSym{0.0}{0.0}
  \end{tikzpicture}.
\end{equation}
As before, we also impose that $C=C^{\dagger}$ and $A^s=(A^s)^{\dagger}$.
For this network, the FPCM proceeds as follows:
\begin{enumerate}
\item Isometrically gauge the uniform MPS composed of tensor $A$.
Using $A$, find the isometric tensor $U$ and (positive) symmetric matrix $C'$ satisfying:
\begin{equation}
  \begin{tikzpicture}[every node/.style={scale=0.90},scale=.55]
   \centerCudtens{0.0}{0.0}{C'}
   \MPSutens{2.0}{0.0}{A}
   \draw (4.0,0.0) node {$\propto$};
   \MPSuLtens{6.0}{0.0}{U}
   \centerCudtens{8.0}{0.0}{C'}
  \end{tikzpicture}.
  \label{eq:fpcm_projector_sym}
\end{equation}
This is performed with a new uniform MPS gauging method described in Appendix~\ref{sec:MPSgauging_sym}.
\label{alg:sym_new_ctmrg_gauge}
\item Obtain the new MPS boundary tensor $A$ and the new CTM $C$ using $U$ found in 
step~\ref{alg:sym_new_ctmrg_gauge}.
This is done by numerically solving the following fixed point equations (in practice using an 
iterative method such as Arnoldi):
\begin{equation}
  \begin{tikzpicture}[every node/.style={scale=1.0},scale=.55]
  \cornerCultens{0.0}{0.0}{C'}
  \cornerIdltens{0.0}{-2.0}
  \MPSutens{2.0}{0.0}{A}
  \MPSdLtens{2.0}{-2.0}{\bar{U}}
  \centerIudtens{4.0}{0.0}
  \cornerIurtens{4.0}{-2.0}
  \draw (6.0,-1.0) node (X) {$\propto$};
  \cornerCultens{8.0}{-1.0}{C'}
  \end{tikzpicture}
  \label{eq:fpcm_renormalize_sym_C}
\end{equation}
\begin{equation}
  \begin{tikzpicture}[every node/.style={scale=0.90},scale=.55]
      \cornerIultens{0.0}{2.0}
      \MPSltens{0.0}{0.0}{A'}
      \cornerIdltens{0.0}{-2.0}
      \MPOtens{2.0}{0.0}{T}
      \MPSuLtens{2.0}{2.0}{U}
      \MPSdLtens{2.0}{-2.0}{\bar{U}}
      \cornerIdrtens{4.0}{2.0}
      \Ilrtens{4.0}{0.0}
      \cornerIurtens{4.0}{-2.0}
      \draw (6.0,0.0) node (X) {$\propto$};
      \MPSltens{8.0}{0.0}{A'}
  \end{tikzpicture}
  \label{eq:fpcm_renormalize_sym}
\end{equation}
(note that Eq.~\ref{eq:fpcm_renormalize_sym_C} may or may not be different from how $C'$ was solved for in
step~\ref{alg:sym_new_ctmrg_gauge} but this alternative fixed point equation may give an improved CTM).
\label{alg:sym_new_ctmrg_boundary}
\end{enumerate}
Then, steps~\ref{alg:sym_new_ctmrg_gauge} and~\ref{alg:sym_new_ctmrg_boundary} are repeated 
until convergence. 
We should point out that the boundary MPS tensor $A$ solved for using the fixed point 
equation Eq.~\eqref{eq:fpcm_renormalize_sym} may only be symmetric up to errors in the accuracy
that the fixed point is solved to, and it may be useful to symmetrize the tensor explicitly
during the optimization.

Note that a similar set of fixed point equations for numerically solving for boundary MPSs
were discussed previously in Ref.~\onlinecite{Haegeman17}.
We would also like to point out that obtaining the isometry as proposed in 
Eq.~\eqref{eq:fpcm_projector_sym} can be viewed as a translationally invariant version of the 
so-called ``simplified one-directional 1D method" discussed in Ref.~\onlinecite{Orus12}.
More generally, the tensor $U$ can be viewed as a translationally invariant version of
the projector obtained in CTMRG.

Note that the following fixed point equation can be used to obtain a more accurate CTM $C$:
\begin{equation}
  \begin{tikzpicture}[every node/.style={scale=0.90},scale=.55]
      \cornerCultens{0.0}{2.0}{C'}
      \MPSltens{0.0}{0.0}{A}
      \cornerIdltens{0.0}{-2.0}
      \MPOtens{2.0}{0.0}{T}
      \MPSutens{2.0}{2.0}{A}
      \MPSdLtens{2.0}{-2.0}{\bar{U}}
      \cornerIurtens{4.0}{2.0}
      \MPSrUtens{4.0}{0.0}{U}
      \cornerIurtens{4.0}{-2.0}
      \cornerIdltens{4.0}{-2.0}
      \draw (6.0,0.0) node (X) {$\propto$};
      \cornerCultens{8.0}{0.0}{C'}
  \end{tikzpicture}.
  \label{eq:fpcm_sym_cornerfixedpoint}
\end{equation}
A similar fixed point equation for the CTM was discussed previously in Ref.~\onlinecite{Vanderstraeten15,
Vanderstraeten16}. 

We also note that in practice, we find performing a few steps of CTMRG per step of the FPCM can 
help improve the convergence of the algorithm and obtain a more accurate fixed point environment.
One can therefore think of FPCM as a way to speed up a CTMRG implementation, by performing
a step of FPCM periodically during the CTMRG algorithm to help speed up convergence.
In that case, it is important to solve for an improved CTM with Eq.~\eqref{eq:fpcm_sym_cornerfixedpoint}
so that the best possible CTM is used for the CTMRG step.

In the next section, we will describe a generalization of this algorithm to asymmetric tensor
networks.

\subsubsection{Asymmetric FPCM}
\label{subsubsec:fpcm_nonsym}

The asymmetric version of FPCM is not as straightforward as the symmetric version, analogous
to the case for CTMRG. Our strategy is to determine translational invariant analogues of the CTMRG
projectors shown in Eq.~\eqref{eq:ctm_CLUrenorm}--\eqref{eq:ctm_CLDrenorm}, and then determine
the environment tensors $\{C_i,A_j\}$ from fixed point equations.

We use the same ansatz as that used for the asymmetric CTMRG algorithm (as presented in 
Sec.~\ref{subsubsec:ctmrg_nonsym}):
\begin{equation}
  \begin{tikzpicture}[every node/.style={scale=0.80},scale=.55]
  \ctmrgansatzTwoTwo{0.0}{0.0}
  \end{tikzpicture}.
  \label{eq:fpcmansatz}
\end{equation}
Using this ansatz for the environment, the left move of FPCM consists of the following steps:
\begin{enumerate}
\item ``Biorthogonalize" the top and bottom MPSs comprised of MPS tensors $A_U$ and $A_D$.
Using $A_{U/D}$, we find $P_L,P_L^-$ along with a new set of $C_{LU}',C_{DL}'$ satisfying:
\begin{equation}
  \begin{tikzpicture}[every node/.style={scale=0.80},scale=.55]
   \centerCudtens{0.0}{0.0}{C_{LU}'}
   \MPSutens{2.0}{0.0}{A_U}
   \draw (4.0,0.0) node {$\propto$};
   \MPSuLtens{6.0}{0.0}{P_L}
   \centerCudtens{8.0}{0.0}{C_{LU}'}
  \end{tikzpicture}
  \label{eq:fpcmleftmove_CLU}
\end{equation}
\begin{equation}
  \begin{tikzpicture}[every node/.style={scale=0.80},scale=.55]
   \centerCudtens{0.0}{0.0}{C_{DL}'}
   \MPSdtens{2.0}{0.0}{A_D}
   \draw (4.0,0.0) node {$\propto$};
   \MPSdLtens{6.0}{0.0}{P_L^-}
   \centerCudtens{8.0}{0.0}{C_{DL}'}
  \end{tikzpicture}
  \label{eq:fpcmleftmove_CLD}
\end{equation}
where $P_L,P_L^-$ satisfy Eq.~\eqref{eq:defPL}.
There are multiple possible methods for finding tensors $P_L,P_L^-$ and $C_{LU}',C_{DL}'$ that
satisfy Eq.~\eqref{eq:fpcmleftmove_CLU}--\eqref{eq:fpcmleftmove_CLD}, and the choices are not unique.
The method we use is described in detail in Appendix~\ref{sec:MPSgauging_nonsym}.
\label{alg:new_corner_gauge_x}
\item Obtain the left HRTM and CTMs using the gauged MPS tensors $P_L,P_L^-$ found in 
step ~\ref{alg:new_corner_gauge_x}.
This is done by numerically solving the following fixed point equations (in practice using an 
iterative method such as Arnoldi):
\begin{equation}
  \begin{tikzpicture}[every node/.style={scale=0.8},scale=.55]
  \cornerCultens{0.0}{0.0}{C_{LU}'}
  \cornerIdltens{0.0}{-2.0}
  \MPSutens{2.0}{0.0}{A_U}
  \MPSdLtens{2.0}{-2.0}{P_L^-}
  \centerIudtens{4.0}{0.0}
  \cornerIurtens{4.0}{-2.0}
  \draw (6.0,-1.0) node (X) {$\propto$};
  \cornerCultens{8.0}{-1.0}{C_{LU}'}
  \end{tikzpicture}
  \label{eq:fpcmleftfp_CLU}
\end{equation}
\begin{equation}
  \begin{tikzpicture}[every node/.style={scale=0.8},scale=.55]
    \ALprenorm{0.0}{0.0}
    \draw (4.5,0.0) node (X) {$\propto$};
    \ALp{6.0}{0.0}.
  \end{tikzpicture}
  \label{eq:fpcmleftmove_AL}
\end{equation}
\begin{equation}
  \begin{tikzpicture}[every node/.style={scale=0.8},scale=.55]
  \cornerIultens{0.0}{0.0}
  \cornerCdltens{0.0}{-2.0}{C_{DL}'}
  \MPSuLtens{2.0}{0.0}{P_L}
  \MPSdtens{2.0}{-2.0}{A_D}
  \cornerIdrtens{4.0}{0.0}
  \centerIudtens{4.0}{-2.0}
  \draw (6.0,-1.0) node (X) {$\propto$};
  \cornerCdltens{8.0}{-1.0}{C_{DL}'}
  \end{tikzpicture}
  \label{eq:fpcmleftfp_CDL}
\end{equation}
(Eq.~\eqref{eq:fpcmleftfp_CLU} and~\eqref{eq:fpcmleftfp_CDL} may be redundant, depending on what
method for obtaining $P_L,P_L^-$ is used).
\label{alg:new_corner_boundary_x}
\end{enumerate}

Steps~\ref{alg:new_corner_gauge_x}--\ref{alg:new_corner_boundary_x} constitute the left move
of the asymmetric FPCM algorithm. 
For a single step of FPCM, the lattice is rotated, and the other directional moves are performed.
For example, one could follow a conventional ordering of the directional CTMRG and next perform 
the up move, then the right move, and then the down move. 
In practice, we don't find that the ordering makes a noticeable difference in the performance
of the algorithm.
Note that a similar set of fixed point equations for the CTMs, HRTMs, and HCTMs were discussed
by Baxter in the context of his CTM method~\cite{Baxter82}.

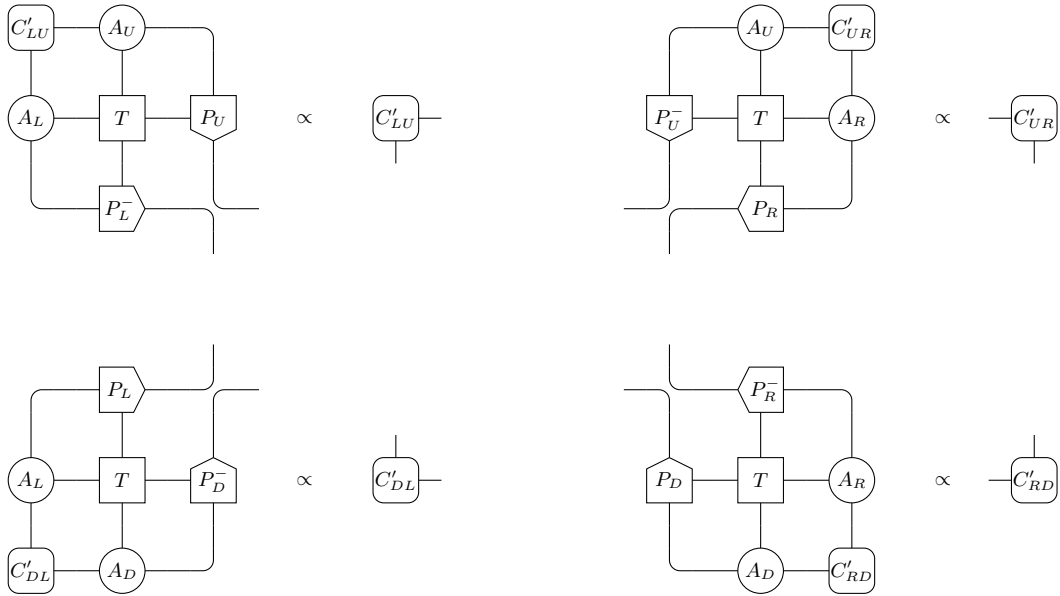
\begin{figure*}[t]
  \begin{tikzpicture}[every node/.style={scale=0.85},scale=.6]
  \def\y{-8.0}
  \def\x{14.0}
  \CULTprenorm{0.0}{0.0}
  \draw[scale=1.0] (4.0,0.0) node (X) {$\propto$};
  \CULp{6.0}{0.0}
  \CDLTprenorm{0.0}{\y}
  \draw[scale=1.0] (4.0,\y) node (X) {$\propto$};
  \CDLp{6.0}{\y}
  \CURTprenorm{\x}{0.0}
  \draw[scale=1.0] (\x+4.0,0.0) node (X) {$\propto$};
  \CURp{\x+6.0}{0.0}
  \CDRTprenorm{\x}{\y}
  \draw[scale=1.0] (\x+4.0,\y) node (X) {$\propto$};
  \CDRp{\x+6.0}{\y}
  \end{tikzpicture}
 \caption{Fixed point equations for the CTMs.
 }
 \label{eq:fpcm_nonsym_cornerfixedpoint}
\end{figure*}

The CTMs can be obtained in an alternative way from the corner transfer fixed 
point equations show in Fig.~\ref{eq:fpcm_nonsym_cornerfixedpoint} (which are generalizations of 
Eq.~\eqref{eq:fpcm_sym_cornerfixedpoint}, the symmetric corner transfer fixed point equation).
We find that obtaining the CTMs with these equations leads to a more accurate environment than 
those found in Eq.~\eqref{eq:fpcmleftfp_CLU} and~\eqref{eq:fpcmleftfp_CDL}, but are of course more
computationally expensive.
Finding more accurate CTMs is important for calculating accurate observables. 
In addition, like in the symmetric FPCM case, we find that in practice the accuracy of this procedure 
is improved by alternating between steps of FPCM and CTMRG, which we will discuss more in 
Section~\ref{sec:results}.
In that case, it is particularly important to get the most accurate CTMs possible by solving for them
with the fixed point equation in Fig.~\ref{eq:fpcm_nonsym_cornerfixedpoint}.

The algorithm looks very similar to the VUMPS algorithm when the network is Hermitian about a 
certain direction (horizontal or vertical), in which case a pair of tensors $P_L,P_L^-$ can be 
chosen to be isometric.
However, like in CTMRG, the CTMs are used explicitly, not the center matrix of 
VUMPS/iDMRG, and the corners can be seen roughly as ``square roots" of the center 
matrix. This is discussed in more detail in Ref.~\onlinecite{Haegeman17}.

The leading cost of this algorithm, the calculation of the new boundaries, is $O(\chi^3 d^2)$ 
where $\chi$ is the bond dimension of the boundary, and $d$ is the bond dimension of the network 
(assuming the fixed point is calculated in a sparse way with an iterative method such as Arnoldi 
and for simplicity assuming a large $\chi$ limit).
This is the same leading cost as single-site VUMPS or single-site iDMRG. 
The cost of CTMRG, following the most standard schemes, is generally a full eigendecomposition, 
singular value decomposition, or QR decomposition of some part of the grown boundary. 
Since the boundary is grown from a bond dimension $\chi$ to a bond dimension $\chi d$, these 
decompositions lead to a scaling of the algorithm of $O(\chi^3 d^3)$, so asymptotically both 
(single site) VUMPS and our new corner method scale better than traditional CTMRG 
in the network bond dimension.

Even so, each step of traditional CTMRG can be much faster than the new schemes presented,
because of the fixed points that we must calculate (note that avoiding the use of fixed point
equations was one of the original motivations for the development of CTMRG as an alternative
to DMRG~\cite{Nishino96,Nishino97}).
However, we will see in Section~\ref{sec:results} that solving for the environment tensors with 
fixed point equations leads to a large speedup in total convergence time, because substantially 
fewer steps are needed for convergence.

The speedup of VUMPS and FPCM over CTMRG is particularly pronounced for networks with small gaps.
One way to understand this is that the original CTMRG can be viewed as a power method, where only 
a single (or pair of) row-to-row and/or column-to-column transfer matrices are absorbed into the 
environment at a time, and the projectors are only determined in a local way.
The new schemes properly exploit the translational invariance of the system, and iterative methods 
such as Arnoldi are known to be much faster than power methods for finding eigenvectors of matrices 
with small gaps (and the gaps of the transfer matrices are expected to be related to the gap of 
the system~\cite{Zauner15}).
In addition, the projectors that are used for renormalization in the FPCM are 
obtained from the current guess for the entire (translationally invariant) boundary, not just a 
set of local tensors.

\section{Results}
\label{sec:results}

Here, we present benchmark results for the methods described in the previous section: CTMRG, transfer
matrix VUMPS, and the new fixed point corner method (FPCM). 
We benchmark the 2D classical ferromagnetic Ising model in Section~\ref{subsec:ising}, the 2D 
classical XY model in Section~\ref{subsec:xy}, the 2D quantum spin-1/2 Heisenberg model in 
Section~\ref{subsec:heisenberg}, and the chiral Resonating Valence Bond (RVB) PEPS in 
Section~\ref{subsec:rvbchiral}.

For all of the examples shown, the networks are on the square lattice and have a single-site unit 
cell, and all tensors used are dense. 
Calculations were performed with a single BLAS thread. 
To obtain a consistent comparison between different methods, the starting boundary states are chosen 
to be small (usually with bond dimension 2), the methods are run until convergence with the small bond 
dimension, and then the bond dimension is increased to the final one (CTMRG is used to grow the 
bond dimension for the FPCM, and the bond dimension growth scheme introduced in 
Ref.~\onlinecite{ZaunerStauber18} is used for VUMPS).
Most of the calculations were performed using the Extreme Science and Engineering Discovery 
Environment (XSEDE)~\cite{XSEDE} with Intel Math Kernel Library (MKL), except calculations
in Fig.~\ref{fig:ising_nonsymmetric}, which were performed on a laptop with OpenBLAS.
Fixed points are calculated using the Arnoldi method as implemented in ARPACK.

\subsection{2D classical Ising model}
\label{subsec:ising}

\begin{figure*}[t]
 \centering
 \subfigure[]
 {
 \includegraphics[width=0.48\linewidth,keepaspectratio=true]{\figpath/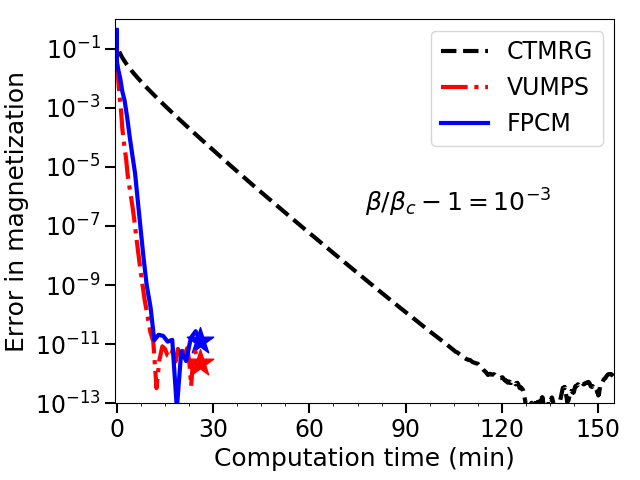}
 \label{fig:ising1a}
 }
 \subfigure[]
 {
 \includegraphics[width=0.48\linewidth,keepaspectratio=true]{\figpath/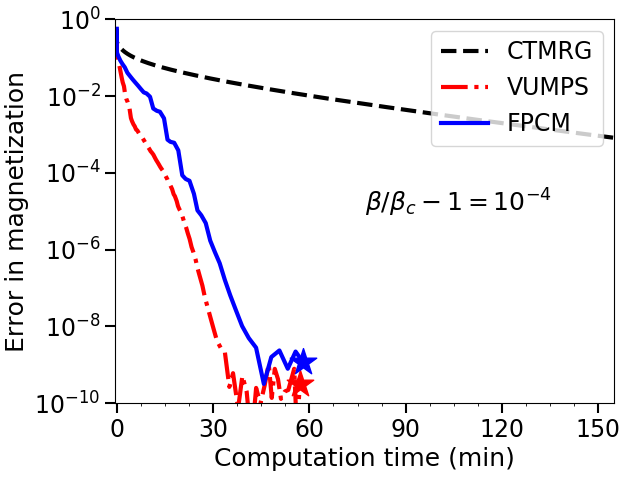}
 \label{fig:ising1b}
 }
 \subfigure[]
 {
 \includegraphics[width=0.6\linewidth,keepaspectratio=true]{\figpath/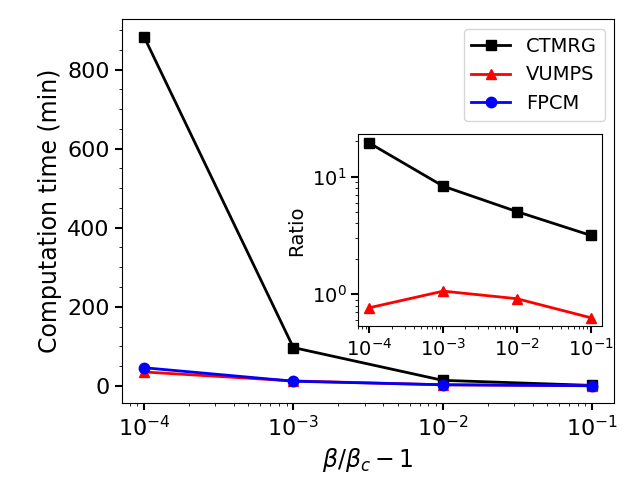}
 \label{fig:ising2}
 }
 \caption{Plots (a) and (b) show the error in the magnetization for the isotropic 2D classical 
Ising model as a function of computation time at two temperatures near criticality, where (b) is 
closer to criticality than (a).
The network has a bond dimension of $d=2$, and a boundary MPS bond dimension of $\chi=600$ is used. 
A fully symmetric CTM ansatz is used for CTMRG and the FPCM, and full symmetry is exploited in 
VUMPS.
The speedup of VUMPS and the corner method over CTMRG increases as one gets closer to criticality.
Stars indicate the environment tensors have reached a fixed point, and data points beyond those 
points are numerical fluctuations and were not shown in order to simplify the plot.
Plot (c) shows convergence time as a function of inverse temperature above criticality, 
$\beta/\beta_c-1$, for the 2D classical Ising model.
For all data points, a boundary MPS bond dimension of $\chi=600$ is used.
All data is converged to an error in the magnetization of $< 2\times 10^{-9}$. 
The inset shows the ratio of the convergence time of CTMRG and VUMPS with respect to the FPCM 
convergence time (note the log scale).
 }
 \label{fig:ising1}
\end{figure*}

In Fig.~\ref{fig:ising1}, we present benchmark results for the isotropic 2D ferromagnetic classical 
Ising model. 
The MPO comprising the partition function for this model has a link bond dimension of $d=2$, and 
the tensor can be taken to be real and symmetric about all rotations and reflections. 
The environment tensors we use for all methods are restricted to being real.
For CTMRG and the FPCM, in the ansatz in Eq.~\eqref{eq:ctmansatz}, we impose 
$A_U=A_R=A_D=A_L\equiv A$ and $C_{LU}=C_{UR}=C_{RD}=C_{DL}\equiv C$, and additionally impose
$A^s=(A^s)^T$ and $C=C^T$. 
For VUMPS, in Eq.~\eqref{eq:vumps}, we impose $[A_U^R]^s=([A_U^L]^s)^T$ and 
$C_U=C_U^T$. 
Additionally, when we calculate observables, we set the bottom fixed point MPS equal to the top 
fixed point MPS.
For CTMRG, we find the projector to renormalize the boundary using a symmetric eigendecomposition, 
which is fast and numerically very stable.

From Fig.~\ref{fig:ising1}, we see that as we approach the critical point of the 2D classical 
Ising model, the performance improvement of VUMPS and our new fixed point corner method (FPCM) over 
CTMRG increases. 
This can be understood by the fact that the boundary tensors for the new methods are obtained
by solving fixed point equations (in practice with Arnoldi and Lanczos methods), which are known 
to be faster than power methods for finding extremal eigenvectors of matrices with small gaps. 
This indicates that these new methods are better suited for studying systems close to or at 
criticality, e.g. in combination with the theory of ``finite entanglement scaling" 
\cite{Tagliacozzo08,Pollmann09,Pirvu12,Stojevic15}.

\begin{figure*}[t]
 \centering
 \subfigure[]
 {
 \includegraphics[width=0.48\linewidth,keepaspectratio=true]{\figpath/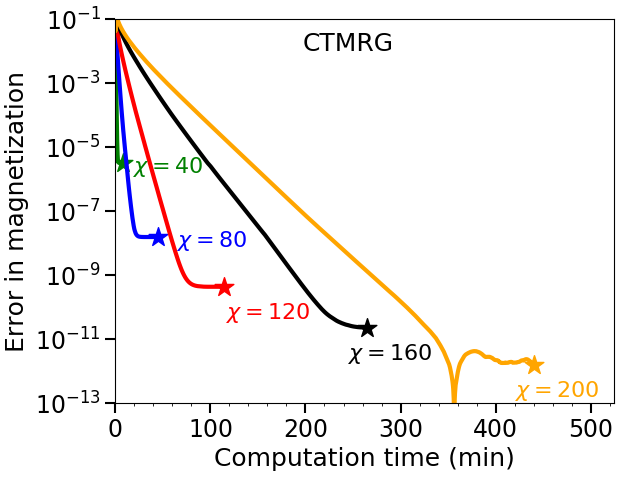}
 \label{subfig:ising_nonsymmetric_corboz}
 }
 \subfigure[]
 {
 \includegraphics[width=0.48\linewidth,keepaspectratio=true]{\figpath/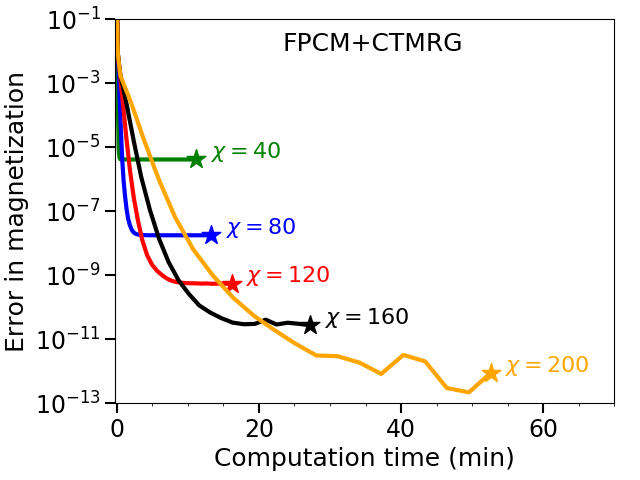}
 \label{subfig:ising_nonsymmetric_fpcm_corboz}
 }
\caption{Plots of error in magnetization for the isotropic ferromagnetic 2D classical Ising 
model at $\beta/\beta_c-1=10^{-3}$ with random non-unitary gauge transformations 
introduced on the horizontal and vertical links, as shown in Eq.~\eqref{eq:randomgauge}.
This artificially breaks the lattice symmetry in order to test each method on an asymmetric 
network. 
Plot (a) shows results for the asymmetric CTMRG algorithm by Corboz et al. in Ref.~\onlinecite{Corboz14} 
and plot (b) shows results for the FPCM introduced in this work combined with the CTMRG algorithm used
in (a).
}
 \label{fig:ising_nonsymmetric}
\end{figure*}

In Fig.~\ref{fig:ising_nonsymmetric}, we present results for the 2D ferromagnetic classical Ising 
model in a non-unitary basis.
In other words, we introduce the following ``gauge transformations" on the links of the tensor 
network: 
\begin{equation}
  \begin{tikzpicture}[every node/.style={scale=0.80},scale=.55]
   \MPOtens{-4.0}{2.0}{T}
   \draw (-2.0,2.0) node {$\to$};
   \gaugeXlrtens{2.0}{4.0}{Y}
   \gaugeXudtens{0.0}{2.0}{X}
   \MPOtens{2.0}{2.0}{T}
   \gaugeXudtens{4.0}{2.0}{X^{-1}}
   \gaugeXlrtens{2.0}{0.0}{Y^{-1}}
  \end{tikzpicture}.
  \label{eq:randomgauge}
\end{equation}
These gauge transformations, for random complex non-unitary matrices $X$ and $Y$, artificially break the 
rotation and reflection symmetries of the Ising model partition function.
Gauge transformations like these can be introduced during a PEPS optimization if explicit 
symmetries are not enforced, even if the state being targeted is expected to be rotationally 
symmetric.
The environments we use for all methods are complex.

In Fig.~\ref{subfig:ising_nonsymmetric_corboz}, we show results for the asymmetric CTMRG method
proposed by Corboz et al. in Ref.~\onlinecite{Corboz14} and reviewed in 
Section~\ref{subsubsec:ctmrg_nonsym}.
In Fig.~\ref{subfig:ising_nonsymmetric_fpcm_corboz}, we show results for performing a step of the
FPCM introduced in Section~\ref{subsubsec:fpcm_nonsym} every five steps of the CTMRG method
of Corboz et al., which leads to substantially improvements in convergence time.
Here, for the FPCM steps we use a pseudoinverse cutoff of $1\times 10^{-7}$ to ensure stability of the
algorithm, which we find is only necessary in early steps when the bond dimension is being increased.
Note that to obtain full precision for the fixed point environments for this example, it was important
to calculate SVD in the biorthogonalization step of CTMRG with the LAPACK routine \lstinline|gesvd| 
as opposed to the routine \lstinline|gesdd|.
We discuss this point and other details about the stability of the CTMRG method for this example in 
Appendix~\ref{sec:CTMRG_asym_alts}.

Note that we found that the asymmetric FPCM method as presented in Section~\ref{subsubsec:fpcm_nonsym}
alone has a tendency to ``get stuck," i.e., not find the most accurate fixed point environment for a given
bond dimension.
It is possible that a modification of the method itself can fix this problem, but we find that
as presented, the method is very simple and numerically stable, and combining with CTMRG
is very effective and robust.
We speculate about the reason why combining FPCM with CTMRG may help improve the accuracy of the
fixed point found with FPCM at the end of Appendix~\ref{sec:MPSgauging_nonsym}.

We also note that the asymmetric versions of CTMRG and FPCM are more computationally demanding and
the convergence in general is not as good as the symmetric versions of the algorithms.
Improving these methods by finding better techniques for obtaining the projectors is an interesting
area of further research.

\subsection{2D classical XY model}
\label{subsec:xy}

\begin{figure*}[t]
 \centering
 \subfigure[]
 {
 \includegraphics[width=0.45\linewidth,keepaspectratio=true]{\figpath/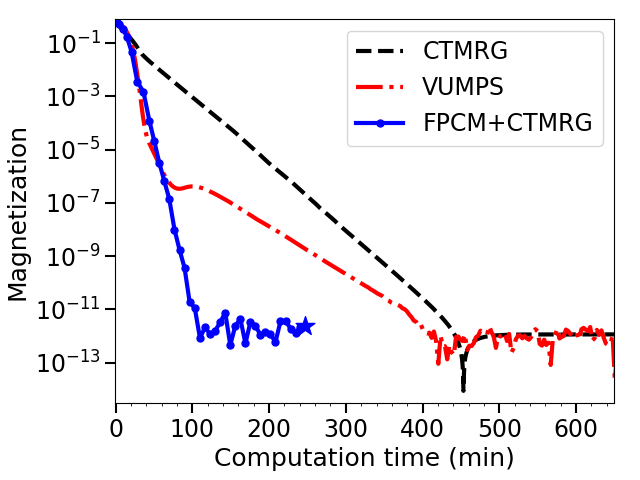}
 \label{fig:xy}
 }
 \subfigure[]
 {
 \includegraphics[width=0.45\linewidth,keepaspectratio=true]{\figpath/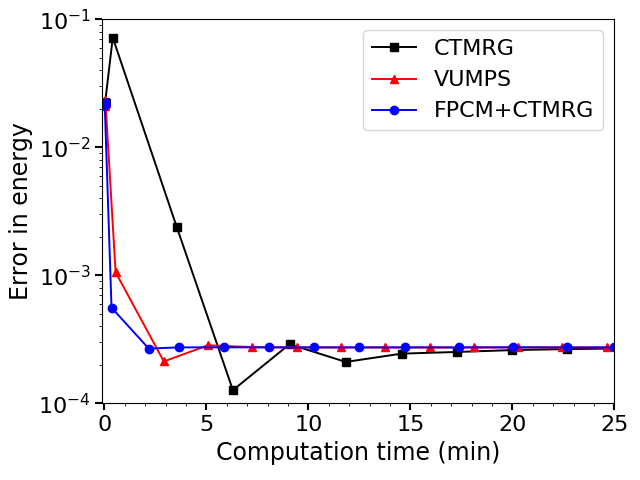}
 \label{fig:heisenberg}
 }
 \subfigure[]
 {
 \includegraphics[width=0.45\linewidth,keepaspectratio=true]{\figpath/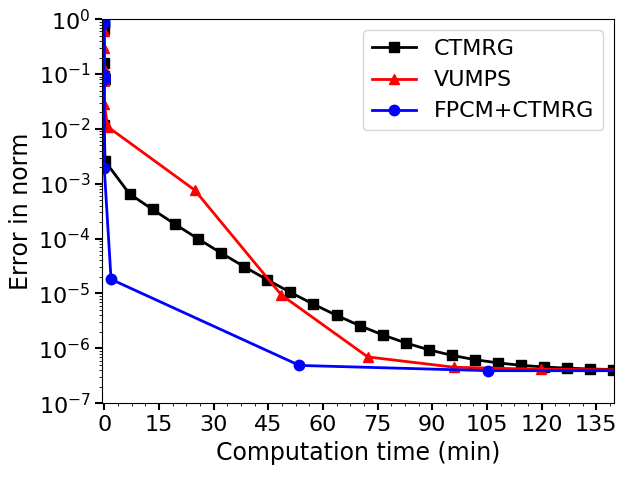}
 \label{fig:rvbchiral}
 }
 \caption{(a) Plot of magnetization for the 2D classical XY model, for network bond dimension 
 $d=25$ and boundary MPS bond dimension $\chi=50$.
 (b) Plot of error in energy (compared to Monte Carlo results) for the 2D quantum Heisenberg model.
 The network bond dimension is $d=25$ (or PEPS bond dimension $D=\sqrt{d}=5$), and the MPS boundary 
 bond dimension $\chi=100$.
 (c) Plot of error in the norm (where the ``exact" results is taken to be an extrapolation of
 the norm in the limit of a large environment bond dimension) of the chiral RVB PEPS.
 The network bond dimension is $d=9$ (or PEPS bond dimension $D=\sqrt{d}=3$), and the boundary MPS 
 bond dimension is $\chi=800$.
 }
\end{figure*}

In Fig.~\ref{fig:xy}, we present results for contracting the partition function for the 2D 
classical XY model.
Because the lattice degree of freedom is continuous for this model, the MPO tensor comprising the 
partition function can only be constructed approximately, though to high accuracy. 
The XY model has been studied previously with transfer matrix DMRG (TMRG)~\cite{Chung99}.
Additionally, related models have been studied previously with CTMRG~\cite{Foster03,Foster03b}.
The construction we use for the partition function is described in Ref.~\onlinecite{XY1,XY2}, 
where HOTRG was used to contract the partition function, and we refer readers to those
references for details on constructing the MPO for this model.
We use an inverse critical temperature $10\%$ below the critical point estimated in that reference,
and use an approximation for the MPO with a link bond dimension of $d=25$.
We use zero applied magnetic field, and at this temperature the model is expected to be gapped.
Since the U(1) symmetry cannot be broken at any finite temperature, we expect the magnetization to 
be zero.

The MPO tensor comprising the partition function is real and symmetric about reflections about the 
diagonals of the network, but not symmetric about the $x$ and $y$ axes.
The environment we use for all methods is restricted to being real.
For CTMRG and the FPCM, in the ansatz in Eq.~\eqref{eq:ctmansatz}, we impose 
$A_U=A_R^T=A_D=A_L^T\equiv A$, $C_{LU}=C_{UR}\equiv C$, and $C_{DL}=C_{RD}\equiv D$, and 
additionally impose $C=C^T$ and $D=D^T$. 
For VUMPS, in Eq.~\eqref{eq:vumps} we don't impose any symmetries, but when we calculate 
observables, we set the bottom fixed point MPS equal to the transpose of the top fixed point MPS 
(such that the environment is invariant under a rotation by $\pi$).

For CTMRG, we obtain the projectors using a symmetric diagonalization of the grown corner.
For FPCM, we obtain the fixed point projectors by isometrically gauging the boundary MPS.
Additionally, like for the asymmetric FPCM calculation performed in Section~\ref{subsec:ising},
we find it is best to alternate between steps of FPCM and CTMRG instead of performing FPCM alone,
and the results shown are obtained by performing a few steps of CTMRG per step of FPCM.
We see that VUMPS performs noticeably worse than the FPCM, likely because the ansatz we use for the
VUMPS calculation does not exploit the lattice symmetry as well as the CTM ansatz.
Like with the Ising model, we expect the improvement of the FPCM compared to CTMRG to become even 
more pronounced closer to the critical point.

\subsection{2D quantum Heisenberg model}
\label{subsec:heisenberg}

In Fig.~\ref{fig:heisenberg}, we present results for contracting a PEPS approximation ($D=\sqrt{d}=5$) 
to the ground state of the 2D quantum Heisenberg model. 
The PEPS tensor was optimized using the conjugate gradient method described in 
Ref.~\onlinecite{Vanderstraeten16}. 
We plot the error in the energy relative to the energy obtained from Monte Carlo 
simulations~\cite{HeisMC}.

The PEPS tensor is complex and symmetric (not Hermitian) about all rotations and reflections,
which was a symmetry imposed in the optimization.
Therefore, the MPO tensor that comprises the tensor network for the norm of the PEPS is also 
complex and symmetric about all rotations and reflections. 
The environments we use for all methods are necessarily complex. 
For CTMRG and the FPCM, in the ansatz in Eq.~\eqref{eq:ctmansatz}, we impose 
$A_U=A_R=A_D=A_L\equiv A$, $C_{LU}=C_{UR}=C_{RD}=C_{DL}\equiv C$, and additionally impose $C=C^T$. 
For VUMPS, in Eq.~\eqref{eq:vumps} we don't impose any symmetries
\footnote{One may expect that we could set $[A_U^R]^s=([A_U^L]^s)^T$ and $C_U=C_U^T$.
However, we were unable to get VUMPS to converge with these constraints imposed,
and it would likely require a nontrivial modifications of the VUMPS algorithm.}, 
but when we calculate observables, we set the bottom fixed point MPS equal to the top fixed point 
MPS (not the conjugate of the top fixed point, as we would do if the MPO was Hermitian as 
opposed to complex symmetric).
The CTMRG algorithm we use is a modification of the one from Ref.~\onlinecite{Corboz14}, where the 
symmetry of the network is exploited wherever possible.
The FPCM method we use is a modification of the asymmetric version presented in 
Section~\ref{subsubsec:fpcm_nonsym} where the symmetry of the network is exploited wherever
possible.
Additionally, we use a modification of the uniform MPS biorthogonalization procedure in 
Appendix~\ref{sec:MPSgauging_nonsym}, where we first gauge the MPSs isometrically before
we biorthogonalize them.
As previously mentioned in Sections~\ref{subsec:ising}--\ref{subsec:xy}, we find for the FPCM that 
it is best to perform a few steps of CTMRG per step of the FPCM, which we find improves the 
accuracy of the fixed point environment.

\subsection{Chiral resonating valence bond PEPS}
\label{subsec:rvbchiral}

In Fig.~\ref{fig:rvbchiral}, we present results for contracting a chiral resonating valence bond 
(RVB) PEPS. 
The chiral RVB PEPS state was introduced as a chiral extension of the traditional nearest neighbor 
RVB PEPS~\cite{chiralRVB1,chiralRVB2}. 
As in the previous works on this model, we choose $\lambda_1 = \lambda_2 = \lambda_{\textrm{chiral}} = 1$,
where $\lambda_2 = \lambda_{\textrm{chiral}} = 0$ would correspond to the non-chiral nearest neighbor 
RVB state.
We refer readers to those previous works on this model for details on its derivation and physics.

The PEPS tensor (and therefore double layer MPO tensor) for this model is complex and Hermitian 
about the horizontal, vertical and diagonal reflections of the lattice.
For CTMRG and the FPCM, in the ansatz in Eq.~\eqref{eq:ctmansatz}, we impose $A_U=A_R=A_D=A_L\equiv A$ 
and $C_{LU}=C_{UR}=C_{RD}=C_{DL}\equiv C$, and additionally impose $A^s=(A^s)^{\dagger}$ and 
$C=C^{\dagger}$. 
For VUMPS, in Eq.~\eqref{eq:vumps} we do not impose any symmetries
\footnote{One may expect that we could set $[A_U^R]^s=([A_U^L]^s)^{\dagger}$ and 
$C_U=C_U^{\dagger}$. 
However, we found in practice these relations only held up to diagonal phases.
This could possibly be fixed by some modification of the VUMPS algorithm.}.
When we calculate observables, we set the bottom fixed point MPS obtained from VUMPS equal to the 
complex conjugate of the top fixed point MPS.
For CTMRG, the projectors are obtained with a Hermitian diagonalization of the grown corner,
and for FPCM, the fixed point projectors are obtained by isometrically gauging the boundary MPS.

Again, we see an improvement in performance of the FPCM and VUMPS over 
CTMRG, but the FPCM performs better than VUMPS (we believe for this case
because the symmetry of the network is exploited better in the CTM ansatz).
Again, we perform a few steps of CTMRG per step of the FPCM, which we find improves the convergence time.

\section{Conclusion and Outlook}
\label{sec:conclusion}

We presented two new approaches for contracting infinite 2D tensor networks, such as 2D classical 
partition functions and 2D quantum states represented as a PEPS.
One approach uses the recently proposed VUMPS algorithm to obtain boundary MPSs that approximate
the infinite environment of the tensor network.
The other approach uses the CTM ansatz like CTMRG, but improves upon CTMRG by
solving for the boundary tensors with fixed point equations, which we refer to as the
fixed point corner method (FPCM).
With careful benchmarking, we compared these new approaches to CTMRG for a variety of systems, 
which is currently the most widely used method for contracting 2D tensor networks in infinite PEPS 
calculations.
We found that both methods improve upon the performance of CTMRG, though for certain models, the 
improvement is more pronounced for FPCM as opposed to VUMPS.

We showed that the improvement upon CTMRG is particularly pronounced 
as models approach criticality, as exemplified by our benchmarking of the 2D classical Ising model. 
This can be explained by the fact that, as the gap of the model closes, so too does the gap of the 
transfer matrix. 
By solving for the boundary tensors with fixed point equations, methods such as Arnoldi and Lanzos 
can be used, which are known to perform better than power methods for finding extremal 
eigenvectors of matrices with small gaps.
Even though each step of the new approaches we present can be slower than each 
step of CTMRG, substantially fewer steps are required to reach fixed points leading to an overall 
improvement in the performance.

We are convinced that these new methods directly improve the performance of current state of the 
art infinite PEPS optimization techniques, where the contraction of the network is the most 
computationally expensive step.
When combined with recently introduced variational methods for optimizing 
PEPS\cite{Corboz16b,Vanderstraeten16}, we expect that significant 
improvements can still be made to existing PEPS algorithms.

\acknowledgments

M.F. acknowledges helpful feedback from P.R. Corboz and T. Nishino. 
M.F. would also like to thank S.R. White and E.M. Stoudenmire for useful input on the presentation
of the results.
The  authors  gratefully  acknowledge  support from the National Science Foundation Graduate 
Research Fellowship Program (NSF GRFP) under Grant No. DGE-1144469 (M.F.), 
the Austrian Science Fund (FWF): F4104 SFB ViCoM and F4014 SFB FoQuS (V.Z.-S. and F.V.), and 
the European Research Council (ERC) under Grant No. 715861 (J.H.).
J.H. and L.V. are supported by the Research Foundation Flanders (FWO).
This project has received funding from the European Research Council (ERC) under the European 
Unions Horizon 2020 research and innovation programme (grant agreement No 647905).
This work used the Extreme Science and Engineering Discovery Environment (XSEDE), which is 
supported by National Science Foundation grant number ACI-1548562.

\appendix

\section{Some comments on the numerical stability of the asymmetric CTMRG method}
\label{sec:CTMRG_asym_alts}

In this section, we discuss the stability of the CTMRG algorithm introduced in Ref.~\onlinecite{Corboz14}
(and reviewed in this work in Section~\ref{subsubsec:ctmrg_nonsym}).
We also discuss slight alternatives to that method and discuss their relevance for the efficiency and
stability of the asymmetric CTMRG algorithm.

We begin by pointing out an equivalent form of the algorithm of Ref.~\onlinecite{Corboz14}:
\begin{enumerate}
 \item We first define the half system transfer matrices $C_U^{(1)}$ and $C_D^{(1)}$:
  \begin{equation}
  \begin{tikzpicture}[every node/.style={scale=0.7},scale=.48]
  \CULT{0.0}{0.0}
  \CURT{2.0}{0.0}
  \CDLT{0.0}{-3.0}
  \CDRT{2.0}{-3.0}
  \draw[scale=1.0] (5.5,-2.0) node (X) {$\equiv$};
  \CudOneOne{10.0}{0.0}
  \end{tikzpicture}
  \label{eq:defCUD}
\end{equation}
   To obtain the projectors, we use the same biorthogonalization procedure as was used in
   Eq.~\eqref{eq:biortho_corboz}--\eqref{eq:biortho_corboz_tildePL}.
   We take the following SVD of $C_D^{(1)} C_U^{(1)}$:
\begin{equation}
  \begin{tikzpicture}[every node/.style={scale=0.95},scale=.55]
  \CUCDrp{0.0}{0.0}
  \draw[scale=1.0] (5.0,0.0) node (X) {$\approx$};
  \svdCUCDrp{7.0}{0.0}
  \end{tikzpicture}.
  \label{eq:biortho_corboz_simplified}
\end{equation}
In other words, we take the SVD $C_D^{(1)} C_U^{(1)} = U_L \Sigma_L^2 V_L^{\dagger}$.
Eq.~\eqref{eq:biortho_corboz_simplified} is approximate because we truncate according to the singular 
values down to the desired bond dimension for the renormalized environment.
\item Now we obtain $P_L,P_L^-$ as follows:
\begin{equation}
  \begin{tikzpicture}[every node/.style={scale=0.95},scale=.55]
  \MPSuLtens{0.0}{0.0}{P_L}
  \draw[scale=1.0] (2.0,0.0) node (X) {$=$};
  \Ilrtens{4.0}{0.5}
  \cornerIultens{4.0}{-0.5}
  \centerCudptens{6.0}{0.0}{C_U^{(1)}}
  \mpsALp{8.0}{0.0}{V_L}
  \centerCudtens{10.0}{0.0}{\Sigma_L^+}
  \end{tikzpicture}
  \label{eq:biortho_corboz_PL_simplified}
\end{equation}
\begin{equation}
  \begin{tikzpicture}[every node/.style={scale=0.95},scale=.55]
  \MPSdLtens{0.0}{0.0}{P_L^-}
  \draw[scale=1.0] (2.0,0.0) node (X) {$=$};
  \Ilrtens{4.0}{-0.5}
  \cornerIdltens{4.0}{0.5}
  \centerCudptens{6.0}{0.0}{C_D^{(1)}}
  \mpsALp{8.0}{0.0}{\bar{U}_L}
  \centerCudtens{10.0}{0.0}{\Sigma_L^+}
  \end{tikzpicture}
  \label{eq:biortho_corboz_tildePL_simplified}
\end{equation}
\end{enumerate}
This alternative for the left move obtains the same tensors $P_L$,$P_L^-$ but skips the step of
calculating the QR decomposition of $C_U^{(1)}$, $C_D^{(1)}$, making it computationally more
efficient.

For both the method of Ref.~\onlinecite{Corboz14} and the equivalent form above, it may be important
to perform a pseudoinverse of the (sqaure root) of the singular values obtained in the biorthogonalization.
Because of the (pseudo)inverse, it is important to calculate the singular values to high accuracy.
However, this can become challenging because to obtain an environment with high accuracy, a large
bond dimension must be used and therefore small singular values will appear.
In practice, we found that a higher precision for the environment could be obtained by calculating 
the SVD with the LAPACK routine \lstinline|gesvd| as opposed to the routine \lstinline|gesdd|.

\begin{figure}
 \centering
 \includegraphics[width=0.95\linewidth,keepaspectratio=true]{\figpath/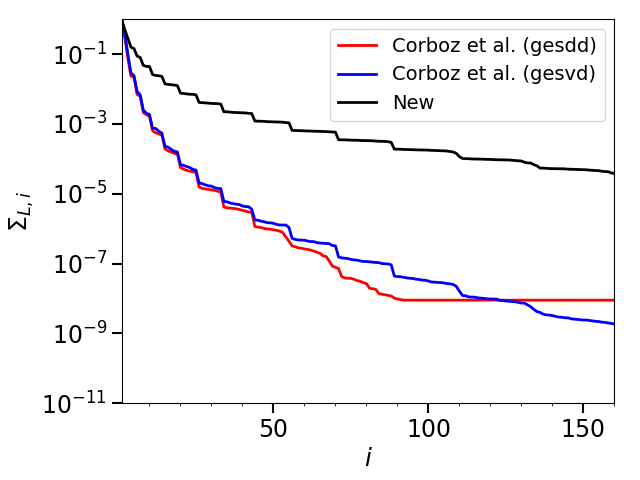}
 \caption{Comparison of spectrums that need to be inverted to perform the biorthogonalization
     in different versions of the asymmetric CTMRG algorithm.
     Note that what is plotted, $\Sigma_{L,i}$, are the square root of the singular values that are 
     calculated during the biorthogonalization. 
     See the main text for details.
 }
 \label{fig:spectrum}
\end{figure}

A comparison of the fixed point spectrum that needs to be inverted during the asymmetric CTMRG 
method of Ref.~\onlinecite{Corboz14} when the SVDs are performed with \lstinline|gesvd| 
and \lstinline|gesdd| is shown in Fig.~\ref{fig:spectrum}.
In that figure, we plot $\Sigma_{L,i}$, the diagonal entries of $\Sigma_L$ 
calculated in Eq.~\eqref{eq:biortho_corboz} or Eq.~\eqref{eq:biortho_corboz_simplified}, for cases when 
CTMRG is run using either \lstinline|gesvd| or \lstinline|gesdd|.
The spectrums are calculated from fixed point environments of the 2D classical Ising model with a 
non-unitary change of basis (as described in Section~\ref{subsec:ising}) with a bond dimension
for the environment of $\chi=160$.
In practice, we found that using \lstinline|gesdd| to calculate the SVD could lead to a less precise
fixed point environment when many small singular values were present, because the small singular
values were not calculated accurately and therefore a pseudoinverse cutoff (i.e. $5\times 10^{-8}$)
was required to ensure the CTMRG algorithm was numerically stable.

Alternative methods can be used to improve the conditioning of the matrix that must be 
inverted during the biorthogonalization procedure.
If the conditioning is improved, the biorthogonalization is less sensitive to the way that the SVD is 
computed.
One method was mentioned in Ref.~\onlinecite{Corboz14}, which is to biorthogonalize the grown CTMs
of the environment (as opposed to the half system transfer matrices).
As mentioned in Ref.~\onlinecite{Corboz14}, this method is not as accurate as using the half system
transfer matrices.
An alternative method to obtain the CTMRG projectors which is as accurate as using the half system
transfer matrices but improves the conditioning of the inverse required in the biorthogonalization 
is as follows:
\begin{enumerate}
\item 
  We start by taking the SVD of the half system transfer matrices:
\begin{equation}
  \begin{tikzpicture}[every node/.style={scale=0.7},scale=.48]
  \CULT{0.0}{0.0}
  \CURT{2.0}{0.0}
  \CDLT{0.0}{-3.0}
  \CDRT{2.0}{-3.0}
  \draw[scale=1.0] (4.5,-2.0) node (X) {$\approx$};
  \SVDU{10.0}{-0.5}
  \SVDD{10.0}{-3.5}
  \draw[scale=1.0] (4.5,-8.5) node (X) {$\approx$};
  \factorU{9.0}{-7.0}
  \factorD{9.0}{-10.0}
  \end{tikzpicture}
  \label{eq:factorize}
\end{equation}
where we define $F_{LU}^s \equiv U_U^s S_U^{1/2}$, $F_{UR}^s \equiv S_U^{1/2} (V_U^s)^{\dagger}$,
$F_{RD}^s \equiv U_D^s S_D^{1/2}$, and $F_{DL}^s \equiv S_D^{1/2} (V_D^s)^{\dagger}$.
Here, one can truncate these equations according to the singular values, which can improve
the computational cost of the algorithm.
\item The next step is to obtain the tensors $P_L,P_L^-$.
Again, we use the same biorthogonalization procedure as was used in
Eq.~\eqref{eq:biortho_corboz}--\eqref{eq:biortho_corboz_tildePL}.
However, this time we biorthogonalize the tensors $F_{LU},F_{DL}$:
\begin{equation}
  \begin{tikzpicture}[every node/.style={scale=0.85},scale=.55]
  \cornerIultens{0.0}{0.0}
  \cornerIdltens{0.0}{-2.0}
  \MPStransferC{2.0}{0.0}{F_{LU}}{F_{DL}}
  \draw[scale=1.0] (4.0,-1.0) node (X) {$\approx$};
  \cornerCultens{6.0}{0.0}{\Sigma_L^2}
  \cornerIdltens{6.0}{-2.0}
  \gaugeARudtens{8.0}{0.0}{\bar{Q}_L}
  \gaugeARudtens{8.0}{-2.0}{W_L}
  \end{tikzpicture}
  \label{eq:biortho_new}
\end{equation}
where the second line is obtained by taking the SVD
$F_{DL}^s F_{LU}^s = W_L \Sigma_L^2 Q_L^{\dagger}$ and truncating according to the singular
values to the desired bond dimension.
\item Again, following the biorthogonalization procedure, we obtain $P_L,P_L^-$ as follows:
\begin{equation}
  \begin{tikzpicture}[every node/.style={scale=0.9},scale=.55]
  \MPSuLtens{0.0}{0.0}{P_L}
  \draw[scale=1.0] (2.0,0.0) node (X) {$=$};
  \MPSuCtens{4.0}{0.0}{F_{LU}}
  \gaugeALudtens{6.0}{0.0}{Q_L}
  \centerCudtens{8.0}{0.0}{\Sigma_L^+}
  \end{tikzpicture}
  \label{eq:biortho_new_PL}
\end{equation}
\begin{equation}
  \begin{tikzpicture}[every node/.style={scale=0.9},scale=.55]
  \MPSdLtens{0.0}{0.0}{P_L^-}
  \draw[scale=1.0] (2.0,0.0) node (X) {$=$};
  \MPSdCtens{4.0}{0.0}{F_{DL}}
  \gaugeALudtens{6.0}{0.0}{\bar{W}_L}
  \centerCudtens{8.0}{0.0}{\Sigma_L^+}
  \end{tikzpicture}
  \label{eq:biortho_new_tildePL}
\end{equation}
For stability of the algorithm, it may be necessary to set a pseudoinverse cutoff 
for $\Sigma_L$, where we use the notation $\Sigma_L^+$ to denote the pseudoinverse.
We found this was particularly relevant in the early steps of the algorithm, during bond 
dimension growth.
\end{enumerate}
We found it could be important to reorthogonalize the tensors $P_L$,$P_L^-$ (i.e. repeat the 
biorthogonalization process on $P_L$,$P_L^-$ a few times) in order to improve the accuracy of the 
projectors obtained.
Note that with this new CTMRG method, it is natural to compute both the left and right projectors 
in a single step to save computation time (by directly biorthogonalizing the tensors $F_{UR}$ and
$F_{RD}$ to obtain the right move projector).

The advantage of this alternative method for obtaining the projectors is that it improves the conditioning
of the inverse required in the biorthogonalization, which in practice means the method is more 
numerically stable (not as sensitive to the choice of SVD algorithm that is used).
The improvement of the conditioning is shown in Fig.~\ref{fig:spectrum}.
The new CTMRG method is more computationally expensive than the one from Ref.~\onlinecite{Corboz14} 
(and the simplified version shown at the beginning of this section) because of the extra SVDs that
are computed.
Alternating between steps of this new asymmetric CTMRG method and steps of FPCM leads to similar speedups
in convergence times as those shown in Section~\ref{subsec:ising}. 

\section{Algorithms for gauging uniform MPSs}

\subsection{New algorithm for isometrically gauging a uniform MPS}
\label{sec:MPSgauging_sym}

Starting with a uniform MPS comprised of the MPS tensor $A$, we would like to find the gauge in 
which the left fixed point of the MPS transfer matrix is identity from bra to ket (the 
``canonical" gauge). 
In other words, we would like to find $U$ and $C$ which satisfy:
\begin{equation}
  \begin{tikzpicture}[every node/.style={scale=1.0},scale=.55]
   \centerCudtens{0.0}{0.0}{C}
   \MPSutens{2.0}{0.0}{A}
   \draw (4.0,0.0) node {$\propto$};
   \MPSuLtens{6.0}{0.0}{U}
   \centerCudtens{8.0}{0.0}{C}
  \end{tikzpicture}
\end{equation}
\begin{equation}
  \begin{tikzpicture}[every node/.style={scale=1.0},scale=.55]
  \defL{0.0}{0.0}{U}{\bar{U}}
  \end{tikzpicture}.
\end{equation}

A method for finding the orthogonal gauge for a uniform MPS was first proposed in the context of the 
iTEBD algorithm\cite{Orus08} and involves a pseudoinverse of the matrix $C$ to solve
for $U$. 
Unfortunately, this means that $U$ is generally only approximately isometric, and the accuracy up 
to which this ``pulling through" equation can be satisfied may be limited for a MPS with small 
singular values.

We now present a fast, robust and highly accurate alternative, where $U$ is constrained
to be isometric and no explicit matrix inversions are used.
Similar to previous methods, we start by finding the left fixed point which we suggestively call 
$C^2$ of the MPS transfer matrix:
\begin{equation}
  \begin{tikzpicture}[every node/.style={scale=1.0},scale=.55]
  \cornerCultens{0.0}{0.0}{C^2}
  \cornerIdltens{0.0}{-2.0}
  \MPSutens{2.0}{0.0}{A}
  \MPSdtens{2.0}{-2.0}{\bar{A}}
  \draw (4.0,-1.0) node (X) {$\propto$};
  \cornerCultens{6.0}{0.0}{C^2}
  \cornerIdltens{6.0}{-2.0}
  \end{tikzpicture}.
\end{equation}

From properties of the transfer matrix, we know that $C^2$ is a positive Hermitian matrix
(up to numerical errors).
We obtain $C$ by taking the square root of $C^2$ (for example by performing a Hermitian
eigendecomposition of $C^2$ and taking the square roots of the positive eigenvalues).
We now obtain our initial $U$ by performing the following polar decomposition: 
\begin{equation}
  \begin{tikzpicture}[every node/.style={scale=1.0},scale=.55]
   \centerCudtens{0.0}{0.0}{C}
   \MPSutens{2.0}{0.0}{A}
   \draw (4.0,0.0) node {$=$};
   \MPSuLtens{6.0}{0.0}{U}
   \centerCudtens{8.0}{0.0}{P}
  \end{tikzpicture}
\label{eq:polardecomp}
\end{equation}
where $U$ is read off as the isometry obtained from the polar decomposition, and $P$ is 
the positive Hermitian matrix obtained from the polar decomposition.
$|C-P|$ is taken to be our initial error in the gauging.
Because we took a square root of $C^2$, our initial error may be limited to the square root of 
machine precision, i.e. $O(10^{-8})$.
If higher precision is required, we repeat the following steps until convergence:
\begin{enumerate}
\item Get a new corner matrix $C$ from the mixed transfer matrix of $A$ and $\bar{U}$ by 
approximately solving for the leading eigenvector of the fixed point equation:
\begin{equation}
  \begin{tikzpicture}[every node/.style={scale=1.0},scale=.55]
  \cornerCultens{0.0}{0.0}{C}
  \cornerIdltens{0.0}{-2.0}
  \MPSutens{2.0}{0.0}{A}
  \MPSdLtens{2.0}{-2.0}{\bar{U}}
  \centerIudtens{4.0}{0.0}
  \cornerIurtens{4.0}{-2.0}
  \draw (6.0,-1.0) node (X) {$\propto$};
  \cornerCultens{8.0}{-1.0}{C}
  \end{tikzpicture}.
\end{equation}
\label{alg:gauge_umps_solve_c}
\item Get a new $U'$ using the new $C$ from step~\ref{alg:gauge_umps_solve_c}.
First, take the (left) polar decomposition of $C$ to get $C=QC'$, where $C'$ is positive and 
Hermitian.
Then, obtain the new $U'$ from a polar decomposition similar to before, i.e.,
\begin{equation}
  \begin{tikzpicture}[every node/.style={scale=1.0},scale=.55]
  \centerCudtens{0.0}{0.0}{C'}
  \MPSutens{2.0}{0.0}{A}
  \draw (4.0,0.0) node {$=$};
  \MPSuLtens{6.0}{0.0}{U'}
  \centerCudtens{8.0}{0.0}{P'}
  \end{tikzpicture}.
\label{eq:polardecomp_prime}
\end{equation}
\label{alg:gauge_umps_solve_u}
\end{enumerate}
We measure the error of the current iteration as $|C'-P'|$, and repeat 
steps~\ref{alg:gauge_umps_solve_c} and \ref{alg:gauge_umps_solve_u} until a desired tolerance 
is met.

\subsection{New algorithm for ``biorthogonalizing" two MPSs}
\label{sec:MPSgauging_nonsym}

We now describe how to ``biorthogonalize" two uniform MPSs (with single-site unit cells) that are 
respectively comprised of MPS tensors $A_U$ and $A_D$.
By biorthogonalize, we mean that we wish to gauge transform $A_U$ and $A_D$ to gauges in which in one 
direction the fixed point of the mixed MPS transfer matrix formed from the two uniform MPSs is the
identity matrix from bra to ket. 
In other words, we would like to gauge transform $A_U$ and $A_D$ so that they satisfy:
\begin{equation}
  \begin{tikzpicture}[every node/.style={scale=0.8},scale=.55]
  \centerCudtens{0.0}{0.0}{C_{LU}}
   \MPSutens{2.0}{0.0}{A_U}
   \draw (4.0,0.0) node {$\propto$};
   \MPSuLtens{6.0}{0.0}{P_L}
   \centerCudtens{8.0}{0.0}{C_{LU}}
  \end{tikzpicture}
\label{eq:pullthroughA}
\end{equation}
\begin{equation}
  \begin{tikzpicture}[every node/.style={scale=0.8},scale=.55]
  \centerCudtens{0.0}{0.0}{C_{DL}}
   \MPSdtens{2.0}{0.0}{A_D}
   \draw (4.0,0.0) node {$\propto$};
   \MPSdLtens{6.0}{0.0}{P_L^-}
   \centerCudtens{8.0}{0.0}{C_{DL}}
  \end{tikzpicture}
\label{eq:pullthroughB}
\end{equation}
where $P_L$ and $P_L^-$ are tensors satisfying
\begin{equation}
  \begin{tikzpicture}[every node/.style={scale=1.0},scale=.55]
  \defPL{0.0}{0.0}
  \end{tikzpicture}.
  \label{eq:defPLfp}
\end{equation}
If $A_D^s=(A_U^s)^{\dagger}$ (possibly after fixing a gauge degree of freedom), one can use the 
approach introduced in Appendix~\ref{sec:MPSgauging_sym}, or use the iTEBD algorithm from 
Ref.~\onlinecite{Orus08}.
In that case, $P_L$ and $P_L^-$ can be chosen to be isometries, such that 
$[P_L^-]^s=(P_L^s)^{\dagger}$ and $(P_L^s)^{\dagger}P_L^s=I$.

If $A_D$ is not the conjugate of $A_U$, then in general $P_L$ and $P_L^-$ won't be isometries, and 
there are multiple possible approaches for satisfying 
Eq.~\eqref{eq:pullthroughA}--\eqref{eq:pullthroughB}.
The approach we use is the following:
\begin{enumerate}
\item We start by getting the left fixed point $C_L$ of the mixed transfer matrix of $A_U$ and $A_D$:
\begin{equation}
\label{eq:fixedpointX2}
  \begin{tikzpicture}[every node/.style={scale=0.8},scale=.55]
      \cornerCultens{0.0}{0.0}{C_L}
      \cornerIdltens{0.0}{-2.0}
      \MPSutens{2.0}{0.0}{A_U}
      \MPSdtens{2.0}{-2.0}{A_D}
      \draw (4.0,-1.0) node (X) {$\propto$};
      \cornerCultens{6.0}{0.0}{C_L}
      \cornerIdltens{6.0}{-2.0}
  \end{tikzpicture}.
\end{equation}
\item We now take the SVD of $C_L = U_L\Sigma_L^2 V_L^{\dagger}$. We define 
$C_{LU}\equiv \Sigma_L V_L^{\dagger}$ and $C_{DL}\equiv U_L\Sigma_L$ and define $P_L,P_L^-$ as follows:
\begin{equation}
  \begin{tikzpicture}[every node/.style={scale=0.8},scale=.55]
   \MPSuLtens{0.0}{0.0}{P_L}
   \draw (2.0,0.0) node {$\equiv$};
   \centerCudtens{4.0}{0.0}{C_{LU}}
   \MPSutens{6.0}{0.0}{A_U}
   \centerCudtens{8.0}{0.0}{C_{LU}^+}
  \end{tikzpicture}
\label{eq:biortho_defPL}
\end{equation}
\begin{equation}
  \begin{tikzpicture}[every node/.style={scale=0.8},scale=.55]
  \MPSdLtens{0.0}{0.0}{P_L^-}
   \draw (2.0,0.0) node {$\equiv$};
   \centerCudtens{4.0}{0.0}{C_{DL}}
   \MPSdtens{6.0}{0.0}{A_D}
   \centerCudtens{8.0}{0.0}{C_{DL}^+}
  \end{tikzpicture}.
  \label{eq:biortho_defPLtilde}
\end{equation}
Here, $C_{LU}^+ = V_L \Sigma_L^+$ and $C_{DL}^+ = \Sigma_L^+ U_L^{\dagger}$, where $\Sigma_L^+$
denotes the pseudoinverse of $\Sigma_L$.
\end{enumerate}

This procedure can be viewed as a simple fixed point formulation of biorthogonalization procedure
introduced by Huang~\cite{Huang11a,Huang11b,Huang12} in the context of non-Hermitian transfer matrix 
DMRG (TMRG) and originally applied to CTMRG by Corboz et al.~\cite{Corboz14}.

In practice, we find that $P_L,P_L^-$ from Eq.~\eqref{eq:pullthroughA}--\eqref{eq:pullthroughB}
may not satisfy Eq.~\eqref{eq:defPLfp} well enough.
If this is the case, we can perform a procedure that we refer to as ``reorthogonalization."
For reorthogonalizing $P_L,P_L^-$ we perform the following steps (which are essentially just the steps
listed above applied to biorthogonalizing $P_L,P_L^-$):
\begin{enumerate}
\item Calculate the dominant left fixed point of the mixed MPS transfer matrix of $P_L,P_L^-$ calculated
from Eq.~\eqref{eq:pullthroughA}--\eqref{eq:pullthroughB} above:
\begin{equation}
\label{eq:projectorid}
  \begin{tikzpicture}[every node/.style={scale=1.0},scale=.55]
    \gaugeXultens{-2.0}{1.0}{Y_L}
    \cornerIdltens{-2.0}{-1.0}
    \MPSuLtens{0.0}{1.0}{P_L}
    \MPSdLtens{0.0}{-1.0}{P_L^-}
    \draw (2.0,0.0) node {$\propto$};
     \gaugeXultens{4.0}{1.0}{Y_L}
     \cornerIdltens{4.0}{-1.0}
   \end{tikzpicture}.
\end{equation}
\item Take the SVD of $Y_L = U\Sigma^2 V^{\dagger}$, defining $Y_{LU}=\Sigma V^{\dagger}$ and 
$Y_{DL}=U \Sigma$. 
Then, update $P_L^s\to Y_{LU} P_L^s Y_{LU}^+$, $[P_L^-]^s\to Y_{DL}^+ [P_L^-]^s Y_{DL}$,
$C_{LU}\to Y_{LU} C_{LU}$, and $C_{DL}\to Y_{DL} C_{DL}$.
\end{enumerate}
These steps can be repeated a number of times. 
Typically, a small number of repetitions (5 to 10) is advantageous for improving the 
accuracy of the biorthogonalization.
It is interesting to point out that a similar concept of reorthogonalization is used in       
standard Krylov subspace methods.

Additionally, we will describe an alternative biorthogonalization method that we have tested with
the asymmetric FPCM. 
One could first gauge transform the MPSs comprised of $A_U$ and $A_D$ isometrically from the left 
(for example using the method described in Appendix~\ref{sec:MPSgauging_sym}) to obtain the isometries which 
we will call $A_U^L$ and $A_D^L$.
Then, the biorthogonalization procedure can be applied to the isometries $A_U^L,A_D^L$ to obtain what 
would in general be a different set of $P_L,P_L^-$ and $C_{LU},C_{DL}$ in 
Eq.~\eqref{eq:biortho_defPL}--\eqref{eq:biortho_defPLtilde}.
Although this seems to improve the conditioning of the inverse, we found that using this method in the FPCM
led to numerical instabilities of the FPCM at larger bond dimensions that were not fixed
by setting a pseudoinverse.
However, we found that this isometric gauging can work when there are symmetry constraints
between the top and bottom MPSs (for example in the Heisenberg model example in 
Section~\ref{subsec:heisenberg}).
A similar instability was noticed by Huang in Ref.~\onlinecite{Huang11a}, where a subspace expansion
technique was proposed in the context of TMRG to improve the stability.
It would be interesting to see if an analgous subspace expansion could be used in the context of FPCM.

In general, it is clear that gauge transforming the MPSs comprised of $A_U$ and $A_D$ before performing
the biorthogonalization can lead to different fixed point projectors $P_L,P_L^-$, which can affect the 
accuracy and stability of FPCM.
As we have mentioned in the main text, in practice we found it worked well to alternate between steps 
of FPCM and CTMRG.
We believe that the steps of CTMRG may help to find ``good" gauges of the HRTMs, HCTMs and CTMs, 
leading  to better projectors obtained from the biorthogonalization.
For a more in-depth discussion of biorthogonalizing uniform MPSs and other possible strategies for 
fixing the gauge, we refer readers to Ref.~\onlinecite{Huang11b}.

\end{document}